# Coupled cavity model for disc-loaded waveguides

M.I. Ayzatsky[1], V.V.Mytrochenko
National Science Center Kharkov Institute of Physics and Technology (NSC KIPT), 610108, Kharkov, Ukraine

The coupled cavity model for calculation characteristics of inhomogeneous disc-loaded waveguides with taking into account the rounding of the disk hole edges was developed. On the base of this model simulation of section tuning process with the field measurement method was conducted. Accuracy of this method under tuning of disc-loaded waveguides with $\beta_{ph} \sim 1$ was evaluated. As it follows from our investigation one can use simple coefficients in tuning process, but the needed values of these coefficients must be obtained by calculation of DLW parameters.

## 1 Introduction

Disc-loaded waveguides (DLW) have been heavily investigated both numerically and analytically over the past seven decades (see, for example, [1,2] and cited there literature). They have also been used, and continue to be used, in a variety of microwave devices such as linear accelerators [3,4], travelling-wave tube amplifiers, backward-wave oscillators [5], etc.

Earlier we have developed approach (a coupled cavity model).that used the eigenmodes of circular cavities as the basic functions for calculation the properties of homogeneous [6] and inhomogeneous DLWs [7,8,9].with cylindrical openings in discs. Here we present the extension of that model for the case of the DLWs with the rounding of the disk hole edges.

## 2 Electromagnetic fields in nonuniform disk-loaded waveguide

Let's consider a cylindrical nonuniform DLW (Fig.1). We will consider only axially symmetric fields with $E_z, E_r, H_\varphi$ components. Time dependence is $\exp(-i\omega t)$ .

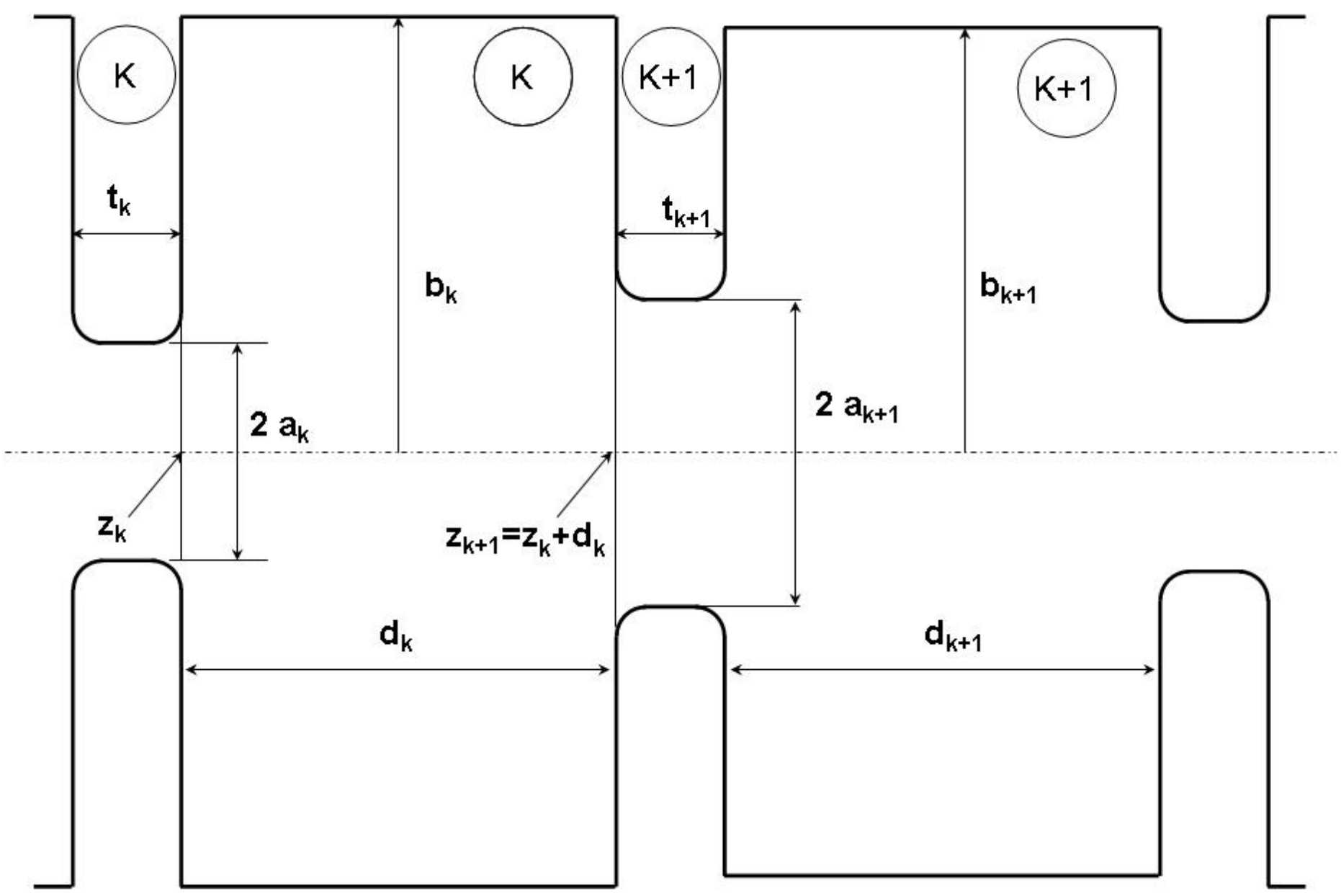

**Fig. 1**

[1] M.I. Aizatskyi, N.I.Aizatsky; aizatsky@kipt.kharkov.ua

We can divide the DLW volume into infinite number of different cylindrical volumes that are contiguous with each other over circle area. In each large volume (cavity) we expand the electromagnetic field with the short-circuit resonant cavity modes

$$\vec{E}^{(k)} = \sum_q e_q^{(k)}(t)\,\vec{E}_q^{(k)}(\vec{r}) \ , \tag{1.1}$$

$$\vec{H}^{(k)} = i\sum_q h_q^{(k)}(t)\,\vec{H}_q^{(k)}(\vec{r})\,, \tag{1.2}$$

where $q = \{0, m, n\}$.

$\vec{E}_q^{(k)}$, $\vec{H}_q^{(k)}$ are the solutions of homogenous Maxwell equations

$$\begin{aligned} rot\,\vec{E}_q^{(k)} &= i\,\omega_q^{(k)}\mu_0\vec{H}_q^{(k)} \ , \\ rot\,\vec{H}_q^{(k)} &= -i\,\omega_q^{(k)}\varepsilon_0\vec{E}_q^{(k)} \end{aligned} \tag{1.3}$$

with boundary condition $\vec{E}_\tau = 0$ on the metal surface

$$E_{0mn,z}^{(k)} = J_0\left(\frac{\lambda_m}{b_k}r\right)\cos\left(\frac{\pi}{d_k}n\left(\,z - z_k\,\right)\right), \tag{1.4}$$

$$H_{0mn,\varphi}^{(k)} = -i\,\omega_{n,m}^{(k)}\frac{\varepsilon_0\,b_k}{\lambda_m}J_1\left(\frac{\lambda_m}{b_k}r\right)\cos\left(\frac{\pi}{d_k}n\left(\,z - z_k\,\right)\right), \tag{1.5}$$

$$E_{0mn,r}^{(k)} = \frac{b_k}{\lambda_m}\frac{\pi\,n}{d_k}J_1\left(\frac{\lambda_m}{b_k}r\right)\sin\left(\frac{\pi}{d_k}n\left(\,z - z_k\,\right)\right), \tag{1.6}$$

$$\omega_{0mn}^{(k)2} = c^2\left\{\left(\frac{\lambda_m}{b_k}\right)^2 + \left(\frac{\pi\,n}{d_k}\right)^2\right\}, \tag{1.7}$$

where $J_0(\lambda_m) = 0$.

Under time dependence $\exp(-i\,\omega t)$ the equations for expansion amplitudes $e_{q,\omega}^{(k)}$ and $h_{q,\omega}^{(k)}$ can be written as

$$\left(\omega_q^{(k)2} - \omega^2\right)e_{q,\omega}^{(k)} = \frac{i\,\omega_q^{(k)}}{N_q^{(k)}}\left(\oint_{S_k}[\vec{E}_c^{(k)}\vec{H}_q^{(k)*}]\,d\vec{S} + \oint_{S_{k+1}}[\vec{E}_c^{(k+1)}\vec{H}_q^{(k)*}]\,d\vec{S}\right), \tag{1.8}$$

$$h_{q,\omega}^{(k)} = -\,i\,\frac{\omega}{\omega_q^{(k)}}e_{q,\omega}^{(k)}\,, \tag{1.9}$$

where $S_k$ is the cross section of the waveguide at the left side of the k-volume,

$$N_{0mn} = \frac{\omega_{0mn,m}^2\,b^4}{c^2\,\lambda_m^2}\varepsilon_0\,\pi\,d\,J_1^2(\lambda_m)\,\bar{\delta}_{n,0}\,, \tag{1.10}$$

$$\bar{\delta}_n = \begin{cases} 1 & n = 0, \\ 0.5 & n > 0. \end{cases}, \tag{1.11}$$

Each small volume inside a disc (waveguide) we divide into a series of homogeneous waveguides (Fig. 2).

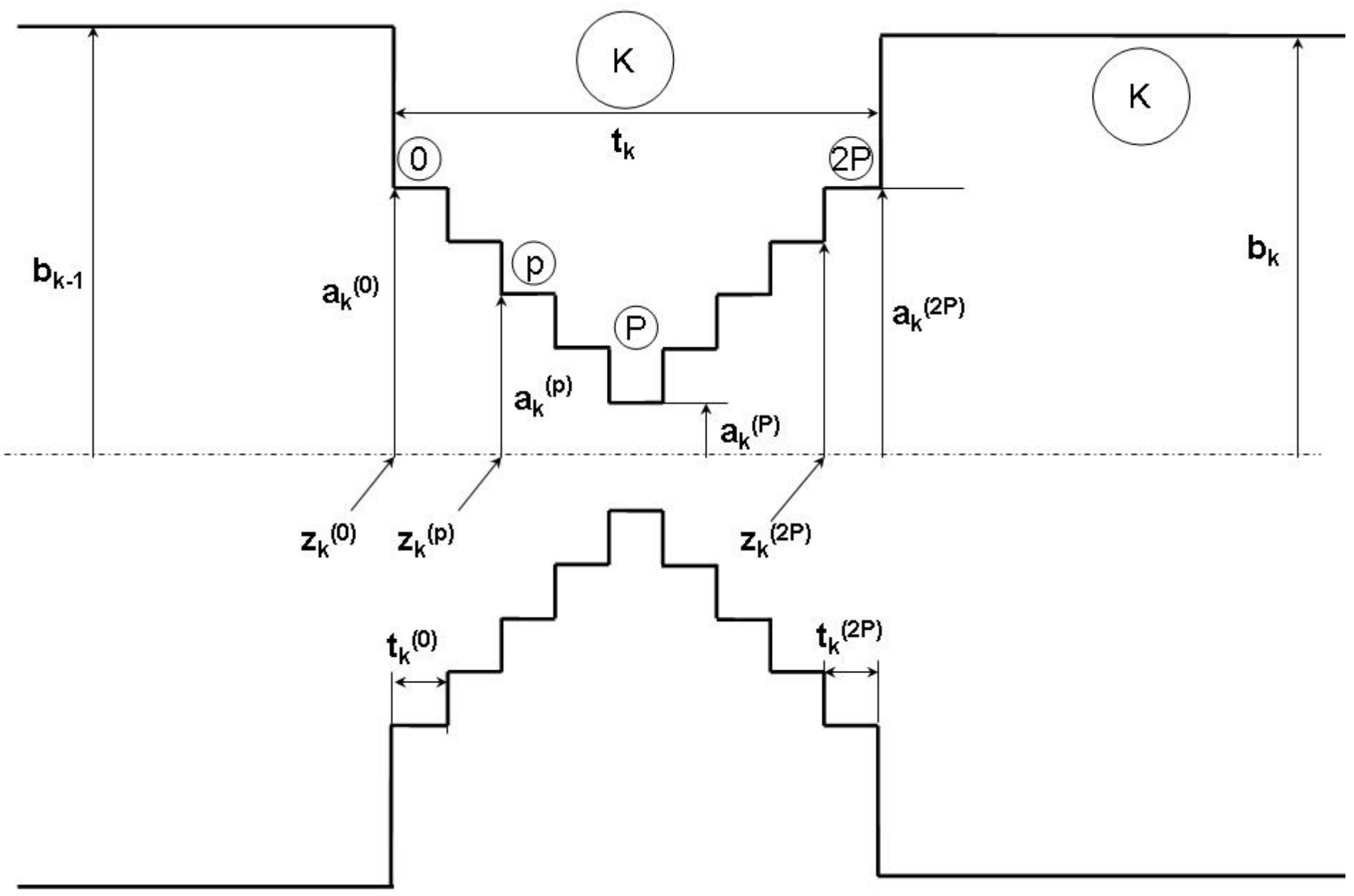

**Fig. 2**

In each homogeneous region we expand the electromagnetic field with the waveguide modes

$$\vec{H}^{(k,p)} = \sum_s \left( C_s^{(k,p)} \vec{\mathcal{H}}_s^{(k,p)} + C_{-s}^{(k,p)} \vec{\mathcal{H}}_{-s}^{(k,p)} \right), \tag{1.12}$$

$$\vec{E}^{(k,p)} = \sum_s \left( C_s^{(k,p)} \vec{\mathcal{E}}_s^{(k,p)} + C_{-s}^{(k,p)} \vec{\mathcal{E}}_{-s}^{(k,p)} \right), \tag{1.13}$$

where

$$\mathcal{E}_{\pm s,z}^{(k,p)} = J_0\left( \frac{\lambda_s}{a_k^{(p)}} r \right) \exp\left\{ \pm \gamma_s^{(k,p)} (z - z_k^{(p)}) \right\}, \tag{1.14}$$

$$\mathcal{H}_{\pm s,\varphi}^{(k,p)} = -i\,\omega \frac{\varepsilon_0\, \varepsilon\, a_k^{(p)}}{\lambda_s} J_1\left( \frac{\lambda_s}{a_k^{(p)}} r \right) \exp\left\{ \pm \gamma_s^{(k,p)} (z - z_k^{(p)}) \right\}, \tag{1.15}$$

$$\mathcal{E}_{\pm s,r}^{(k,p)} = \mp \frac{a_k^{(p)}}{\lambda_s} \gamma_s^{(k,p)} J_1\left( \frac{\lambda_s}{a_k^{(p)}} r \right) \exp\left\{ \pm \gamma_s^{(k)} (z - z_k^{(p)}) \right\}, \tag{1.16}$$

$$\gamma_s^{(k,p)2} = \left( \frac{\lambda_s}{a_k^{(p)}} \right)^2 - \frac{\omega^2}{c^2} = \frac{1}{a_k^{(p)2}} \left( \lambda_s^2 - \frac{a_k^{(p)2} \omega^2}{c^2} \right) = \frac{1}{a_k^{(p)2}} \left( \lambda_s^2 - \Omega_k^{(p)2} \right) = \frac{1}{a_k^{(p)2}} \overline{\gamma}_s^{(k,p)2}, \tag{1.17}$$

$$\Omega_k^{(p)} = \frac{a_k^{(p)} \omega}{c} = \frac{a_k^{(p)} 2\pi f}{c}, \tag{1.18}$$

$$p = 0, 1, \ldots, 2P. \tag{1.19}$$

Magnetic field boundary conditions must be satisfied at the interface between the cavities and waveguides:

$$\sum_s \left( C_s^{(k,0)} \mathcal{H}_{s,\varphi}^{(k,0)} + C_{-s}^{(k)} \mathcal{H}_{-s,\varphi}^{(k)} \right) = i \sum_q h_q^{(k-1)} H_{q,\varphi}^{(k-1)} \tag{1.20}$$

at the plane located at $z = z_{k-1} + d_{k-1}$, and

$$\sum_s \left( C_s^{(k,0)} \mathcal{H}_{s,\varphi}^{(k,0)} + C_{-s}^{(k)} \mathcal{H}_{-s,\varphi}^{(k)} \right) = i \sum_q h_q^{(k)} H_{q,\varphi}^{(k)} \tag{1.21}$$

at $z = z_k$.

Since the magnetic waveguide modes are orthogonal, from the previous equations we can obtain

$$\left\{C_s^{(k,0)} + C_{-s}^{(k,0)}\right\} = \sum_{n,m} (-1)^n e_{0mn}^{(k-1)} \Gamma_{m,s}^{(k,k-1)} \ , \tag{1.22}$$

$$\left\{\exp\left(\gamma^{(k,2P)} t_k^{(2P)}\right) C_s^{(k,2P)} + \exp\left(-\gamma^{(k,2P)} t_k^{(2P)}\right) C_{-s}^{(k,2P)}\right\} = \sum_{n,m} e_{0mn}^{(k)} \Gamma_m^{(k,k)} \ , \tag{1.23}$$

where

$$\Gamma_{m,s}^{(k,k)} = -\frac{2\lambda_s}{J_1(\lambda_s)} F_s\left(\frac{a_k^{(2P)} \lambda_m}{b_k}\right), \tag{1.24}$$

$$\Gamma_{m,s}^{(k,k-1)} = -\frac{2\lambda_s}{J_1(\lambda_s)} F_s\left(\frac{a_k^{(0)} \lambda_m}{b_{k-1}}\right), \tag{1.25}$$

$$F_s(x) = \frac{J_0(x)}{(x)^2 - (\lambda_s)^2}. \tag{1.26}$$

The equations (1.8) we can rewrite as

$$\begin{gathered}\left(\omega_{0mn}^{(k)2} - \omega^2\right) e_{0mn}^{(k)} = -\frac{2\omega_{0m0}^{(k)2}}{b_k^2 d_k J_1^2(\lambda_m) \bar{\delta}_{n,0}} \times \\ \times\left[\sum_{s'} \frac{J_1(\lambda_{s'})}{\lambda_{s'}} \left\{\bar{\gamma}_{s'}^{(k,2P)} \left(a_k^{(2P)}\right)^3 F_{s'}\left(\frac{a_k^{(2P)} \lambda_m}{b_k}\right) C_{r,s'}^{(k)} - (-1)^n \bar{\gamma}_{s'}^{(k+1,0)} \left(a_{k+1}^{(0)}\right)^3 F_{s'}\left(\frac{a_{k+1}^{(0)} \lambda_m}{b_k}\right) C_{l,s'}^{(k+1)}\right\}\right],\end{gathered} \tag{1.27}$$

where

$$\begin{aligned} C_{r,s}^{(k)} &= \left[-C_s^{(k,2P)} \exp\left(\gamma_s^{(k,2P)} t_k^{(2P)}\right) + C_{-s}^{(k,2P)} \exp\left(-\gamma_s^{(k,2P)} t_k^{(2P)}\right)\right] \\ C_{l,s}^{(k)} &= \left[-C_s^{(k,0)} + C_{-s}^{(k,0)}\right] \end{aligned} \tag{1.28}$$

Indexes $l$ (left) and $r$ (right) refer to the coefficients on the left and right sides of the disk.

We will suppose that for $1 \le p \le P$: $a_k^{(p)} > a_k^{(p+1)}$ and for $P+1 \le p \le 2P$: $a_k^{(p)} < a_k^{(p+1)}$.

The magnetic field boundary conditions inside a disc at $z = z_k^{(p)}$ are

$$\sum_s \left(C_s^{(k,p-1)} \mathcal{H}_{s,\varphi}^{(k,p-1)} + C_{-s}^{(k,p-1)} \mathcal{H}_{-s,\varphi}^{(k,p-1)}\right) = \sum_s \left(C_s^{(k,p)} \mathcal{H}_{s,\varphi}^{(k,p)} + C_{-s}^{(k,p)} \mathcal{H}_{-s,\varphi}^{(k,p)}\right), \tag{1.29}$$

$1 \le p \le P$,

$$0 \le r < a_k^{(p)} \ , \tag{1.30}$$

$P+1 \le p \le 2P$,

$$0 \le r < a_k^{(p-1)} \ . \tag{1.31}$$

The electric field boundary conditions inside a disc at $z = z_k^{(p)}$ are

$1 \le p \le P$,

$$\sum_s \left(C_s^{(k,p-1)} \mathcal{E}_{s,r}^{(k,p-1)} + C_{-s}^{(k,p-1)} \mathcal{E}_{-s,r}^{(k,p-1)}\right) = \begin{cases} 0, & a_k^{(p)} \le r < a_k^{(p-1)} \\ \sum_s \left(C_s^{(k,p)} \mathcal{E}_{s,r}^{(k,p)} + C_{-s}^{(k,p)} \mathcal{E}_{-s,r}^{(k,p)}\right), & r < a_k^{(p)} \end{cases} , \tag{1.32}$$

$P+1 \le p \le 2P$,

$$\sum_s \left(C_s^{(k,p)} \mathcal{E}_{s,r}^{(k,p)} + C_{-s}^{(k,p)} \mathcal{E}_{-s,r}^{(k,p)}\right) = \begin{cases} 0, & a_k^{(p-1)} \le r < a_k^{(p)} \\ \sum_s \left(C_s^{(k,p-1)} \mathcal{E}_{s,r}^{(k,p-1)} + C_{-s}^{(k,p-1)} \mathcal{E}_{-s,r}^{(k,p-1)}\right), & r < a_k^{(p-1)} \end{cases}. \tag{1.33}$$

We arrange the amplitudes $C_{\pm s}^{(k,p)}$ ( $s=1,2,3....\infty$ ) in the column vectors $C_{\pm}^{(k,p)}$

$$C_{\pm}^{(k,p)}=\begin{pmatrix} C_{\pm1}^{(k,p)} \\ C_{\pm2}^{(k,p)} \\ C_{\pm3}^{(k,p)} \\ .... \end{pmatrix}, \tag{1.34}$$

Using the orthogonal properties of electric field we transform equalities (1.29),(1.32), (1.33) into matrix ones

$1\le p\le P$

$$\left[C_{+}^{(k,p)}+C_{-}^{(k,p)}\right]=H\left(a_k^{(p-1)},a_k^{(p)}\right)\left[\exp\left\{\gamma^{(k,p-1)}t_k^{(p-1)}\right\}C_{+}^{(k,p-1)}+\exp\left\{-\gamma^{(k,p-1)}t_k^{(p-1)}\right\}C_{-}^{(k,p-1)}\right], \tag{1.35}$$

$$\left[\exp\left\{\gamma^{(k,p-1)}t_k^{(p-1)}\right\}C_{+}^{(k,p-1)}-\exp\left\{-\gamma^{(k,p-1)}t_k^{(p-1)}\right\}C_{-}^{(k,p-1)}\right]=E\left(a_k^{(p-1)},a_k^{(p)}\right)\left[C_{+}^{(k,p)}-C_{-}^{(k,p)}\right], \tag{1.36}$$

$P+1\le p\le 2P$

$$\left[\exp\left\{\gamma^{(k,p-1)}t_k^{(p-1)}\right\}C_{+}^{(k,p-1)}+\exp\left\{-\gamma^{(k,p-1)}t_k^{(p-1)}\right\}C_{-}^{(k,p-1)}\right]=H\left(a_k^{(p)},a_k^{(p-1)}\right)\left[C_{+}^{(k,p)}+C_{-}^{(k,p)}\right], \tag{1.37}$$

$$\left[C_{+}^{(k,p)}-C_{-}^{(k,p)}\right]=E(a_k^{(p)},a_k^{(p-1)})\left[\exp\left\{\gamma^{(k,p-1)}t_k^{(p-1)}\right\}C_{+}^{(k,p-1)}-\exp\left\{-\gamma^{(k,p-1)}t_k^{(p-1)}\right\}C_{-}^{(k,p-1)}\right], \tag{1.38}$$

where the elements of matrices $H$ , $E$ and $\exp\left\{\pm\gamma^{(k,p)}t_k^{(p)}\right\}$ are

$$H_{s,s'}\left(a_k^{(p)},a_k^{(p-1)}\right)=-\frac{2\lambda_s}{J_1\left(\lambda_s\right)}\left(\frac{a_k^{(p)}}{a_k^{(p-1)}\lambda_{s'}}\right)^3 F_s\left(\frac{a_k^{(p-1)}\lambda_{s'}}{a_k^{(p)}}\right), \tag{1.39}$$

$$E_{s,s'}(a_k^{(p)},a_k^{(p-1)})=-\frac{a_k^{(p-1)}}{a_k^{(p)}}\frac{2J_1\left(\lambda_{s'}\right)\gamma_{s'}^{(k,p-1)}}{J_1^2\left(\lambda_s\right)\lambda_{s'}\lambda_s\gamma_s^{(k,p)}}F_{s'}\left(\frac{a_k^{(p-1)}\lambda_s}{a_k^{(p)}}\right), \tag{1.40}$$

$$\left[\exp\left\{\pm\gamma^{(k,p)}t_k^{(p)}\right\}\right]_{s,s'}=\delta_{s,s'}\exp\left\{\pm\gamma_s^{(k,p)}t_k^{(p)}\right\}. \tag{1.41}$$

The matrix equalities (1.35),(1.37),(1.38) we can rewrite using the Generalized Scattering Matrix (GSM) of the waveguide step

$1\le p\le P$

$$\begin{bmatrix}\exp\left\{\gamma^{(k,p-1)}t_k^{(p-1)}\right\}C_{+}^{(k,p-1)} \\ C_{-}^{(k,p)}\end{bmatrix}=\begin{bmatrix}S_{11}^{(k,p-1:p,l)} & S_{12}^{(k,p-1:p,l)} \\ S_{21}^{(k,p-1:p,)} & S_{22}^{(k,p-1:p,l)}\end{bmatrix}\begin{bmatrix}\exp\left\{-\gamma^{(k,p-1)}t_k^{(p-1)}\right\}C_{-}^{(k,p-1)} \\ C_{+}^{(k,p)}\end{bmatrix}, \tag{1.42}$$

$$\begin{bmatrix}S_{11}^{(k,p-1:p,l)} & S_{12}^{(k,p-1:p,l)} \\ S_{21}^{(k,p-1:p,)} & S_{22}^{(k,p-1:p,l)}\end{bmatrix}==\begin{bmatrix}-I+2H^{-1}\left(E+H^{-1}\right)^{-1} & H^{-1}+H^{-1}\left(E+H^{-1}\right)^{-1}\left(E-H^{-1}\right) \\ 2\left(E+H^{-1}\right)^{-1} & \left(E+H^{-1}\right)^{-1}\left(E-H^{-1}\right)\end{bmatrix}, \tag{1.43}$$

$$\begin{aligned} H&=H\left(a_k^{(p-1)},a_k^{(p)}\right) \\ E&=E\left(a_k^{(p-1)},a_k^{(p)}\right)\end{aligned}, \tag{1.44}$$

$P+1\le p\le 2P$

$$\begin{bmatrix}\exp\left\{\gamma^{(k,p-1)}t_k^{(p-1)}\right\}C_{+}^{(k,p-1)} \\ C_{-}^{(k,p)}\end{bmatrix}=\begin{bmatrix}S_{11}^{(k,p-1:p,r)} & S_{12}^{(k,p-1:p,r)} \\ S_{21}^{(k,p-1:p,r)} & S_{22}^{(k,p-1:p,r)}\end{bmatrix}\begin{bmatrix}\exp\left\{-\gamma^{(k,p-1)}t_k^{(p-1)}\right\}C_{-}^{(k,p-1)} \\ C_{+}^{(k,p)}\end{bmatrix}, \tag{1.45}$$

$$\begin{bmatrix} S_{11}^{(k,p-1:p,r)} & S_{12}^{(k,p-1:p,r)} \\ S_{21}^{(k,p-1:p,r)} & S_{22}^{(k,p-1:p,r)} \end{bmatrix} = \begin{bmatrix} -I + 2H\left(I+EH\right)^{-1}E & H + H\left(I+EH\right)^{-1}(I-EH) \\ 2\left(I+EH\right)^{-1}E & \left(I+EH\right)^{-1}\left(I-EH\right) \end{bmatrix}, \quad (1.46)$$

$$\begin{aligned} H &= \widehat{H}\left(a_k^{(p)}, a_k^{(p-1)}\right) \\ E &= \widehat{E}(a_k^{(p)}, a_k^{(p-1)}) \end{aligned}. \quad (1.47)$$

Here, $I$ is a unit matrix and $(\ )^{-1}$ represents the inverse matrix.

Then the GSM of the whole aperture

$$\begin{bmatrix} C_{+}^{(k,0)} \\ C_{-}^{(k,p)}\exp\left\{-\gamma^{(k,2P)}t_k^{(2P)}\right\} \end{bmatrix} = \begin{bmatrix} S_{11}^{(k)} & S_{12}^{(k)} \\ S_{21}^{(k)} & S_{22}^{(k)} \end{bmatrix} \begin{bmatrix} C_{-}^{(k,0)} \\ C_{+}^{(k,2P)}\exp\left\{\gamma^{(k,2P)}t_k^{(2P)}\right\} \end{bmatrix} \quad (1.48)$$

we can find as

$$\begin{aligned} &S^{(k)} = Sw^{(k,0)} \otimes S^{(k,0,1,l)} \otimes Sw^{(k,1)} \otimes S^{(k,1,2,l)} \otimes Sw^{(k,2)} ............ \otimes S^{(k,P-1,P,l)} \otimes Sw^{(k,P)} \otimes \\ &S^{(k,P,P+1,r)} \otimes Sw^{(k,P+1)} ........ \otimes S^{(k,2P-1,2P,r)} \otimes Sw^{(k,2P)} \end{aligned}, \quad (1.49)$$

where the star product $S^{(c)} = S^{(b)} \otimes S^{(a)}$ is determined as follows:

$$\begin{aligned} S_{11}^{(c)} &= \left[ S_{11}^{(a)} + S_{12}^{(a)} S_{11}^{(b)} \left(I - S_{22}^{(a)} S_{11}^{(b)}\right)^{-1} S_{21}^{(a)} \right] \\ S_{12}^{(c)} &= S_{12}^{(a)} \left[ S_{11}^{(b)} \left(I - S_{22}^{(a)} S_{11}^{(b)}\right)^{-1} S_{22}^{(a)} S_{12}^{(b)} + S_{12}^{(b)} \right] \\ S_{21}^{(c)} &= S_{21}^{(b)} \left(I - S_{22}^{(a)} S_{11}^{(b)}\right)^{-1} S_{21}^{(a)} \\ S_{22}^{(c)} &= \left[ S_{21}^{(b)} \left(I - S_{22}^{(a)} S_{11}^{(b)}\right)^{-1} S_{22}^{(a)} S_{12}^{(b)} + S_{22}^{(b)} \right] \end{aligned}. \quad (1.50)$$

Matrix $Sw^{(k,p)}$ is the $S$ matrix of the uniform section of waveguide

$$\begin{bmatrix} Sw_{11}^{(k,p)} & Sw_{12}^{(k,p)} \\ Sw_{21}^{w(k,p)} & Sw_{22}^{(k,p)} \end{bmatrix} = \begin{bmatrix} 0 & \exp\left\{-\gamma^{(k,p)}t_k^{(p)}\right\} \\ \exp\left\{-\gamma^{(k,p)}t_k^{(p)}\right\} & 0 \end{bmatrix}. \quad (1.51)$$

Using(1.48), from (1.22) and (1.23) we can find the amplitudes in the end waveguides and auxiliary coefficients (1.28)

$$\begin{aligned} C_{r,s}^{(k)} &= \sum_{n,m} e_{0mn}^{(k)} \Xi_{m,s}^{(2P,k,k)} + \sum_{n,m} (-1)^n e_{0mn}^{(k-1)} \Xi_{m,s}^{(2P,k,k-1)} \\ C_{l,s}^{(k)} &= \sum_{n,m} e_{0mn}^{(k)} \Xi_{m,s}^{(0,k.k)} + \sum_{n,m} (-1)^n e_{0mn}^{(k-1)} \Xi_{m,s}^{(0,k.k-1)} \end{aligned}, \quad (1.52)$$

where

$$\begin{aligned} \Xi_{m,s}^{(0,k,k)} &= \left[\Theta_{21}^{(k)} - S_{11}^{(k)}\Theta_{21}^{(k)} - S_{12}^{(k)}\Theta_{11}^{(k)}\right]\Gamma_m^{(k,k)} \\ \Xi_{m,s}^{(0,k,k-1)} &= \left[\Theta_{22}^{(k)} - S_{11}^{(k)}\Theta_{22}^{(k)} - S_{12}^{(k)}\Theta_{12}^{(k)}\right]\Gamma_m^{(k,k-1)} \\ \Xi_{m,s}^{(2P,k,k)} &= \left[S_{21}^{(k)}\Theta_{21}^{(k)} + S_{22}^{(k)}\Theta_{11}^{(k)} - \Theta_{11}^{(k)}\right]\Gamma_m^{(k,k)} \\ \Xi_{m,s}^{(2P,k,k-1)} &= \left[S_{21}^{(k)}\Theta_{22}^{(k)} + S_{22}^{(k)}\Theta_{12}^{(k)} - \Theta_{12}^{(k)}\right]\Gamma_m^{(k,k-1)} \end{aligned}, \quad (1.53)$$

$$\begin{bmatrix} \Theta_{11}^{(k)} & \Theta_{12}^{(k)} \\ \Theta_{21}^{(k)} & \Theta_{22}^{(k)} \end{bmatrix} = \begin{bmatrix} (I+S_{22}^{(k)}) & S_{21}^{(k)} \\ S_{12}^{(k)} & (I+S_{11}^{(k)}) \end{bmatrix}^{-1}$$

$$\Theta_{11}^{(k)} = (I+S_{22}^{(k)})^{-1} + (I+S_{22}^{(k)})^{-1} S_{21}^{(k)} \left[ (I+S_{11}^{(k)}) - S_{12}^{(k)} (I+S_{22}^{(k)})^{-1} S_{21}^{(k)} \right]^{-1} S_{12}^{(k)} (I+S_{22}^{(k)})^{-1}$$

$$\Theta_{21}^{(k)} = -\left[ (I+S_{11}^{(k)}) - S_{12}^{(k)} (I+S_{22}^{(k)})^{-1} S_{21}^{(k)} \right]^{-1} S_{12}^{(k)} (I+S_{22}^{(k)})^{-1} \quad . (1.54)$$

$$\Theta_{12}^{(k)} = -(I+S_{22}^{(k)})^{-1} S_{21}^{(k)} \left[ (I+S_{11}^{(k)}) - S_{12}^{(k)} (I+S_{22}^{(k)})^{-1} S_{21}^{(k)} \right]^{-1}$$

$$\Theta_{22}^{(k)} = \left[ (I+S_{11}^{(k)}) - S_{12}^{(k)} (I+S_{22}^{(k)})^{-1} S_{21}^{(k)} \right]^{-1}$$

Lets us chose the $E_{010}$ modes of the cavities as the basic modes (oscillators) [6,10]. In this case the equations (1.52) we can write in the form

$$\begin{aligned} &C_{r,s}^{(k)} - \sum_{n,m}{}' e_{0mn}^{(k)} \Xi_{m,s}^{(2P,k,k)} - \sum_{n,m}{}' (-1)^n e_{0mn}^{(k-1)} \Xi_{m,s}^{(2P,k,k-1)} = e_{010}^{(k)} \Xi_{1,s}^{(2P,k,k)} + e_{010}^{(k-1)} \Xi_{1,s}^{(2P,k,k-1)} \\ &C_{l,s}^{(k)} - \sum_{n,m}{}' e_{0mn}^{(k)} \Xi_{m,s}^{(0,k.k)} - \sum_{n,m}{}' (-1)^n e_{0mn}^{(k-1)} \Xi_{m,s}^{(0,k.k-1)} = e_{010}^{(k)} \Xi_{1,s}^{(0,k.k)} + e_{010}^{(k-1)} \Xi_{1,s}^{(0,k.k-1)} \end{aligned} \quad (1.55)$$

where $\sum_{n,m}{}' = \sum_{n,m;n\neq 0,m\neq 1}$ .

Substituting (1.27) into (1.55) gives a set of equations

$$\begin{aligned} &\tilde{C}_{l,s}^{(k)} + \sum_{s'} \tilde{C}_{r,s'}^{(k)} T_{l,s,s'}^{(k,1)} + \sum_{s'} \tilde{C}_{l,s'}^{(k)} T_{l,s,s'}^{(k,2)} - \sum_{s'} \tilde{C}_{l,s'}^{(k+1)} T_{l,s,s'}^{(k,3)} - \sum_{s'} \tilde{C}_{r,s'}^{(k-1)} T_{l,s,s'}^{(k,4)} = \\ &= R_{l,s}^{(k,1)} e_{010}^{(k)} - R_{l,s}^{(k,2)} e_{010}^{(k-1)} = \sum_j \delta_{j,k} \left\{ R_s^{(k,1)} e_{010}^{(j)} - R_s^{(k,2)} e_{010}^{(j-1)} \right\} = \sum_j \left\{ \delta_{j,k} R_{l,s}^{(k,1)} - \delta_{j+1,k} R_{l,s}^{(k,2)} \right\} e_{010}^{(j)} \\ &\tilde{C}_{r,s}^{(k)} + \sum_{s'} \tilde{C}_{r,s'}^{(k)} T_{r,s,s'}^{(k,1)} + \sum_{s'} \tilde{C}_{l,s'}^{(k)} T_{r,s,s'}^{(k,2)} - \sum_{s'} \tilde{C}_{l,s'}^{(k+1)} T_{r,s,s'}^{(k,3)} - \sum_{s'} \tilde{C}_{r,s'}^{(k-1)} T_{r,s,s'}^{(k,4)} = \\ &= R_{r,s}^{(k,1)} e_{010}^{(k)} - R_{r,s}^{(k,2)} e_{010}^{(k-1)} = \sum_j \delta_{j,k} \left\{ R_{r,s}^{(k,1)} e_{010}^{(j)} - R_{r,s}^{(k,2)} e_{010}^{(j-1)} \right\} = \sum_j \left\{ \delta_{j,k} R_{r,s}^{(k,1)} - \delta_{j-1,k} R_{r,s}^{(k,2)} \right\} e_{010}^{(j)} \end{aligned} \quad ,(1.56)$$

where

$$\begin{aligned} \tilde{C}_{r,s'}^{(k)} &= \frac{\overline{\gamma}_{s'}^{(k,2P)}}{\lambda_{s'}} C_{r,s'}^{(k)} J_1(\lambda_{s'}) \\ \tilde{C}_{l,s'}^{(k)} &= \frac{\overline{\gamma}_{s'}^{(k,0)}}{\lambda_{s'}} C_{l,s'}^{(k)} J_1(\lambda_{s'}) \end{aligned} \quad , \quad (1.57)$$

$$\begin{aligned} T_{l,s,s'}^{(k,1)} &= \frac{\overline{\gamma}_s^{(k,0)} J_1(\lambda_s)}{\lambda_s} \frac{2a_k^{(2P)3}}{b_k^3} \sum_m \Lambda_m^{(1,k)} \frac{\lambda_m^2}{J_1^2(\lambda_m)} F_{s'}\left( \frac{a_k^{(2P)} \lambda_m}{b_k} \right) \Xi_{m,s}^{(0,k,k)} \\ T_{l,s,s'}^{(k,3)} &= \frac{\overline{\gamma}_s^{(k,0)} J_1(\lambda_s)}{\lambda_s} \frac{2a_{k+1}^{(0)3}}{b_k^3} \sum_m \Lambda_m^{(2,k)} \frac{\lambda_m^2}{J_1^2(\lambda_m)} F_{s'}\left( \frac{a_{k+1}^{(0)} \lambda_m}{b_k} \right) \Xi_{m,s}^{(0,k,k)} \\ T_{l,s,s'}^{(k,4)} &= -\frac{\overline{\gamma}_s^{(k,0)} J_1(\lambda_s)}{\lambda_s} \frac{2a_{k-1}^{(2P)3}}{b_{k-1}^3} \sum_m \Lambda_m^{(2,k-1)} \frac{\lambda_m^2}{J_1^2(\lambda_m)} F_{s'}\left( \frac{a_{k-1}^{(2P)} \lambda_m}{b_{k-1}} \right) \Xi_{m,s}^{(0,k,k-1)} \\ T_{l,s,s'}^{(k,2)} &= -\frac{\overline{\gamma}_s^{(k,0)} J_1(\lambda_s)}{\lambda_s} \frac{2a_k^{(0)3}}{b_{k-1}^3} \sum_{n,m} \Lambda_m^{(1,k-1)} \frac{\lambda_m^2}{J_1^2(\lambda_m)} F_{s'}\left( \frac{a_k^{(0)} \lambda_m}{b_{k-1}} \right) \Xi_{m,s}^{(0,k,k-1)} \end{aligned} \quad , \quad (1.58)$$

$$R_{l,s}^{(k,1)}=\frac{\overline{\gamma}_s^{(k,0)}J_1\left(\lambda_s\right)}{\lambda_s}\Xi_{1,s}^{(0,k,k)}$$
$$R_{l,s}^{(k,2)}=-\frac{\overline{\gamma}_s^{(k,0)}J_1\left(\lambda_s\right)}{\lambda_s}\Xi_{1,s}^{(0,k,k-1)},\qquad(1.59)$$

$$T_{r,s,s'}^{(k,1)}=\frac{\overline{\gamma}_s^{(k,2P)}J_1\left(\lambda_s\right)}{\lambda_s}\frac{2a_k^{(2P)3}}{b_k^3}\sum_m\frac{\Lambda_m^{(1,k)}\lambda_m^2}{J_1^2(\lambda_m)}F_{s'}\left(\frac{a_k^{(2P)}\lambda_m}{b_k}\right)\Xi_{m,s}^{(2P,k,k)}$$
$$T_{r,s,s'}^{(k,3)}=\frac{\overline{\gamma}_s^{(k,2P)}J_1\left(\lambda_s\right)}{\lambda_s}\frac{2a_{k+1}^{(0)3}}{b_k^3}\sum_m\frac{\Lambda_m^{(2,k)}\lambda_m^2}{J_1^2(\lambda_m)}F_{s'}\left(\frac{a_{k+1}^{(0)}\lambda_m}{b_k}\right)\Xi_{m,s}^{(2P,k,k)}$$
$$T_{r,s,s'}^{(k,4)}=-\frac{\overline{\gamma}_s^{(k,2P)}J_1\left(\lambda_s\right)}{\lambda_s}\frac{2a_{k-1}^{(2P)3}}{b_{k-1}^3}\sum_m\frac{\Lambda_m^{(2,k-1)}\lambda_m^2}{J_1^2(\lambda_m)}F_{s'}\left(\frac{a_{k-1}^{(2P)}\lambda_m}{b_{k-1}}\right)\Xi_{m,s}^{(2P,k,k-1)},\qquad(1.60)$$
$$T_{r,s,s'}^{(k,2)}=-\frac{\overline{\gamma}_s^{(k,2P)}J_1\left(\lambda_s\right)}{\lambda_s}\frac{2a_k^{(0)3}}{b_{k-1}^3}\sum_m\frac{\Lambda_m^{(1,k-1)}\lambda_m^2}{J_1^2(\lambda_m)}F_{s'}\left(\frac{a_k^{(0)}\lambda_m}{b_{k-1}}\right)\Xi_{m,s}^{(2P,k,k-1)}$$

$$R_{r,s}^{(k,1)}=\frac{\overline{\gamma}_s^{(k,2P)}J_1\left(\lambda_s\right)}{\lambda_s}\Xi_{1,s}^{(2P,k,k)}$$
$$R_{r,s}^{(k,2)}=-\frac{\overline{\gamma}_s^{(k,2P)}J_1\left(\lambda_s\right)}{\lambda_s}\Xi_{1,s}^{(2P,k,k-1)},\qquad(1.61)$$

$$\Lambda_m^{(1,k)}=\begin{cases}\dfrac{1}{b_kh_1^{(k)}}\left[cth(d_kh_1^{(k)})-\dfrac{1}{d_kh_1^{(k)}}\right], & m=1\\[2ex] \dfrac{1}{b_kh_m^{(k)}}cth(d_kh_m^{(k)}), & m>1\end{cases},\qquad(1.62)$$

$$\Lambda_m^{(2,k)}=\begin{cases}\dfrac{1}{b_kh_1^{(k)}}\left[\dfrac{1}{sh(d_kh_1^{(k)})}-\dfrac{1}{d_kh_1^{(k)}}\right], & m=1\\[2ex] \dfrac{1}{b_kh_m^{(k)}sh(d_kh_m^{(k)})}, & m>1\end{cases},\qquad(1.63)$$

$$h_m^{(k)}=\sqrt{\frac{\lambda_m^2}{b_k^2}-\frac{\omega^2}{c^2}}\,.\qquad(1.64).$$

The systems (1.56) are the infinite linear coupled systems of equations, so one can represent its solution in the form

$$\tilde{C}_{l,s}^{(k)}=\sum_j\left(Y_{l,s}^{(1,k,j)}-Y_{l,s}^{(2,k,j)}\right)e_{010}^{(j)}$$
$$\tilde{C}_{r,s}^{(k)}=\sum_j\left(Y_{r,s}^{(1,k,j)}-Y_{r,s}^{(2,k,j)}\right)e_{010}^{(j)},\qquad(1.65)$$

where the new unknowns $Y_{l,s}^{(1,k,j)},Y_{l,s}^{(2,k,j)},Y_{r,s}^{(1,k,j)},Y_{r,s}^{(2,k,j)}$ are the solutions of such sets[2] of equations

[2] Instead two sets of unknowns we introduce four sets of unknowns, so we can add two new sets of equations

$$Y_{l,s}^{(1,k+1,j)}+\sum_{s'}Y_{r,s'}^{(1,k+1,j)}T_{l,s,s'}^{(k+1,1)}+\sum_{s'}Y_{l,s'}^{(1,k+1,j)}T_{l,s,s'}^{(k+1,2)}-\sum_{s'}Y_{l,s'}^{(1,k+2,j)}T_{l,s,s'}^{(k+1,3)}-\sum_{s'}Y_{r,s'}^{(1,k,j)}T_{l,s,s'}^{(k+1,4)}$$
$$=R_{l,s}^{(k+1,1)}\delta_{k+1,j}$$
$$Y_{r,s}^{(1,k,j)}+\sum_{s'}Y_{r,s'}^{(1,k,j)}T_{r,s,s'}^{(k,1)}+\sum_{s'}Y_{l,s'}^{(1,k,j)}T_{r,s,s'}^{(k,2)}-\sum_{s'}Y_{l,s'}^{(1,k+1,j)}T_{r,s,s'}^{(k,3)}-\sum_{s'}Y_{r,s'}^{(1,k-1,j)}T_{r,s,s'}^{(k,4)}= \quad , \tag{1.66}$$
$$=R_{r,s}^{(k,1)}\delta_{k,j}$$

$$Y_{l,s}^{(2,k+1,j)}+\sum_{s'}Y_{r,s'}^{(2,k+1,j)}T_{l,s,s'}^{(k+1,1)}+\sum_{s'}Y_{l,s'}^{(2,k+1,j)}T_{l,s,s'}^{(k+1,2)}-\sum_{s'}Y_{l,s'}^{(2,k+2,j)}T_{l,s,s'}^{(k+1,3)}-\sum_{s'}Y_{r,s'}^{(2,k,j)}T_{l,s,s'}^{(k+1,4)}=$$
$$=R_{l,s}^{(k+1,2)}\delta_{k,j} \quad . \tag{1.67}$$
$$Y_{r,s}^{(2,k,j)}+\sum_{s'}Y_{r,s'}^{(2,k,j)}T_{r,s,s'}^{(k,1)}+\sum_{s'}Y_{l,s'}^{(2,k,j)}T_{r,s,s'}^{(k,2)}-\sum_{s'}Y_{l,s'}^{(2,k+1,j)}T_{r,s,s'}^{(k,3)}-\sum_{s'}Y_{r,s'}^{(2,k-1,j)}T_{r,s,s'}^{(k,4)}=$$
$$=R_{r,s}^{(k,2)}\delta_{k,j+1}$$

Using (1.65) we can rewrite (1.27) as

$$\left(\omega_{0mn}^{(k)2}-\omega^2\right)e_{0mn}^{(k)}=\omega_{0mn}^{(k)2}\sum_{j=-\infty}^{\infty}e_{010}^{(j)}\alpha_{mn}^{(k,j)} , \tag{1.68}$$

where

$$\alpha_{mn}^{(k,j)}=\frac{2\pi\varepsilon_0 b_k^3}{N_{m,n}^{(k)}\lambda_m^3}\left[-\sum_s\left(Y_{r,s}^{(1,k,j)}-Y_{r,s}^{(2,k,j)}\right)F_s\left(\frac{a_k}{b_k}\lambda_m\right)+\right.$$
$$\left.+(-1)^n\sum_s\left(Y_{l,s}^{(1,k+1,j)}-Y_{l,s}^{(2,k+1,j)}\right)F_s\left(\frac{a_{k+1}}{b_k}\lambda_m\right)\right] . \tag{1.69}$$

Let’s note that relations (1.69) are the exact ones.

It follows from (1.68) that for finding the amplitudes of the main $E_{010}$ mode we have to solve a system of coupled equations

$$\left(\omega_{010}^{(k)2}-\omega^2\right)e_{010}^{(k)}=\omega_{010}^{(k)2}\sum_{j=-\infty}^{\infty}e_{010}^{(j)}\alpha_{10}^{(k,j)} . \tag{1.70}$$

Amplitudes of other modes ( $(m,n)\neq(1,0)$ ) can be found by summing the relevant series

$$e_{0mn}^{(k)}=\frac{\omega_{0mn}^{(k)2}}{\left\{\omega_{0mn}^{(k)2}\left(1-\alpha_{m,n}^{(k,k)}\right)-\omega^2\right\}}\sum_{j=-\infty}^{\infty}e_{010}^{(j)}\alpha_{mn}^{(k,j)} . \tag{1.71}$$

The system of coupled equations (1.70) is very similar to that one that can be constructed on the base of equivalent circuits approach. But in our approach the coefficients $\alpha_{m,n}^{(k,j)}$ are electrodynamically strictly defined and can be calculated with necessary accuracy. They depend on both the frequency $\omega$ and geometrical sizes of all volumes.

As it follows from physics of couplings, the contribution of ”long range” couplings is small[3] and one can confine oneself to consideration a finite number of adjacent couplings. In this case the system of coupled equations (1.70) transforms into such one

$$\left(\omega_{010}^{(k)2}-\omega^2\right)e_{010}^{(k)}=\omega_{010}^{(k)2}\sum_{j=k-N}^{k+N}e_{010}^{(j)}\alpha_{10}^{(k,j)} . \tag{1.72}$$

For dealing with this system we have to develop a procedure for $\alpha_{10}^{(k,j)}$ calculation. From (1.69) it follows that we have to know the coefficients

$$\left(Y_{l,s}^{(1,k+1,j)},Y_{r,s}^{(1,k,j)}\right),\ \left(Y_{l,s}^{(2,k+1,j)},Y_{r,s}^{(2,k,j)}\right) \tag{1.73}$$

---

[3] In the case under consideration (when $E_{010}$ modes are the basic ones) ”long range” coupling is realized through evanescent fields

for $k-N \le j \le k+N$. Coefficients $Y$ are the solutions of the infinite set of infinite systems of linear equations. We suppose that this set can be truncated and the solution of such truncated set of equations converge under increasing the number of equations. Besides, we have to make restriction on the number of waveguide modes ($M$) in the field expansion inside apertures and on the number of radial modes ($L$) in the field expansion inside resonators. Used truncation method is explained in [7]. In addition to the case without rounding we have additional parameter $2P+1$ - number of homogeneous waveguides that represent the real opening in the disk.

It is necessary to introduce some criterion for choosing the values of $M, L, P$ that can give correct simulation results. As such criterion we chose the frequency calculation accuracy for the given phase shift per cell in the case of homogeneous DLW. Results of our previous investigations have shown [7,9] that $N \le 4$ (number of "side couplings") gives good results in all cases that are interesting for accelerator physics. The amplitudes $e_{010}^{(k)}$ of homogeneous DLW obey the Floquet theorem

$$e_{010}^{(k)} = e_{010}^{(0)} \exp(ik\varphi) \tag{1.74}$$

and the dispersive equation has such form [6,7,9]

$$\left\{\omega_{010}^{(0)2}\left(1+\alpha_{1,0}^{(0,0)}\right)-\omega^2\right\} = 2\omega_{010}^{(0)2}\sum_{j=1}^{N}\alpha_{10}^{(0,j)}(\omega)\cos(j\varphi). \tag{1.75}$$

For given $\varphi = \varphi_*, N$ and geometrical sizes we can find from this equation corresponding frequency $\omega = \omega_*$.

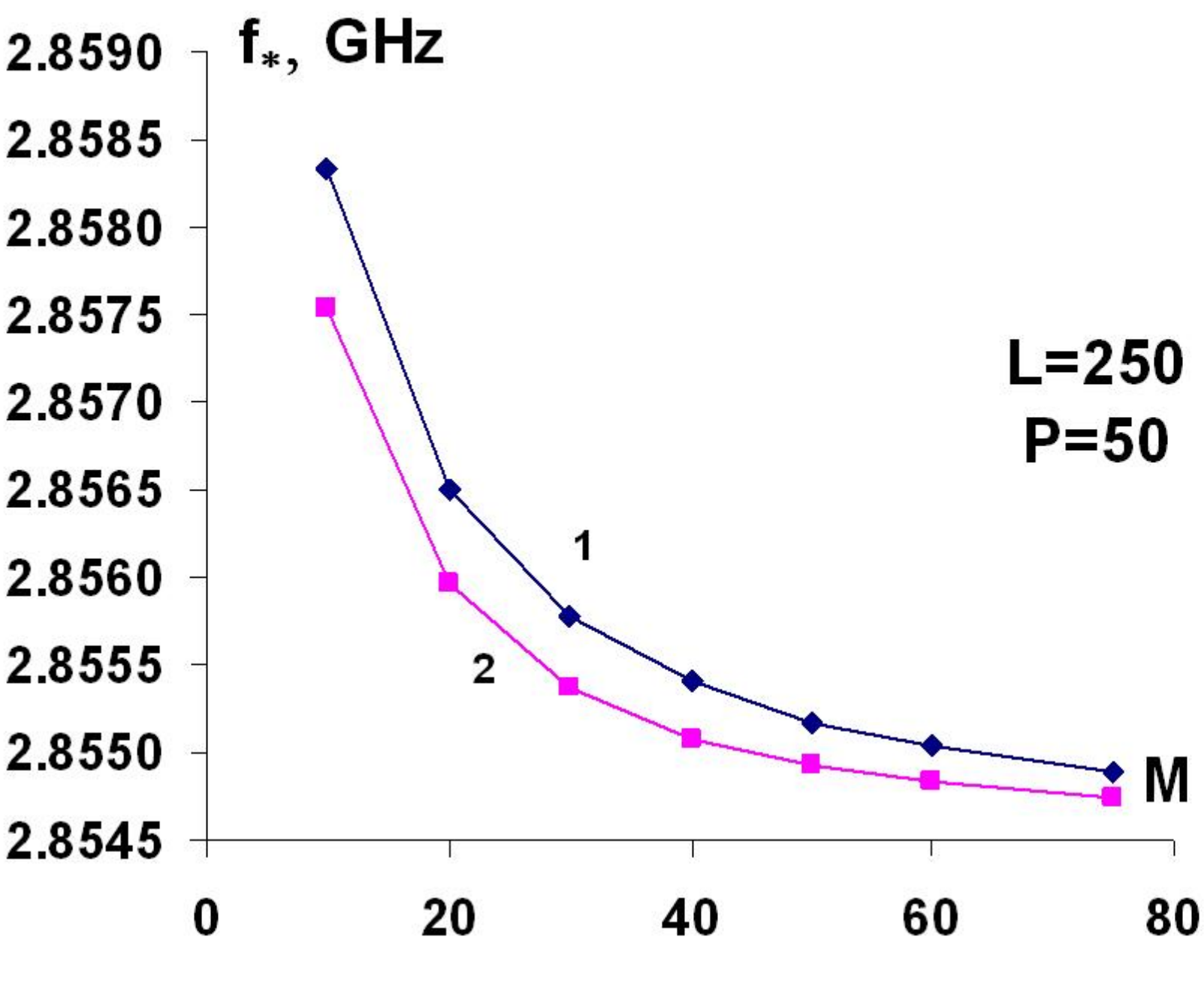

**Fig. 3 Dependence of frequency of $2\pi/3$ mode on number of expansion modes M for homogeneous DLW with $d=$3.099 cm, $t=$0.4 cm, $b=$4.134 cm, $a=$1.2 cm**

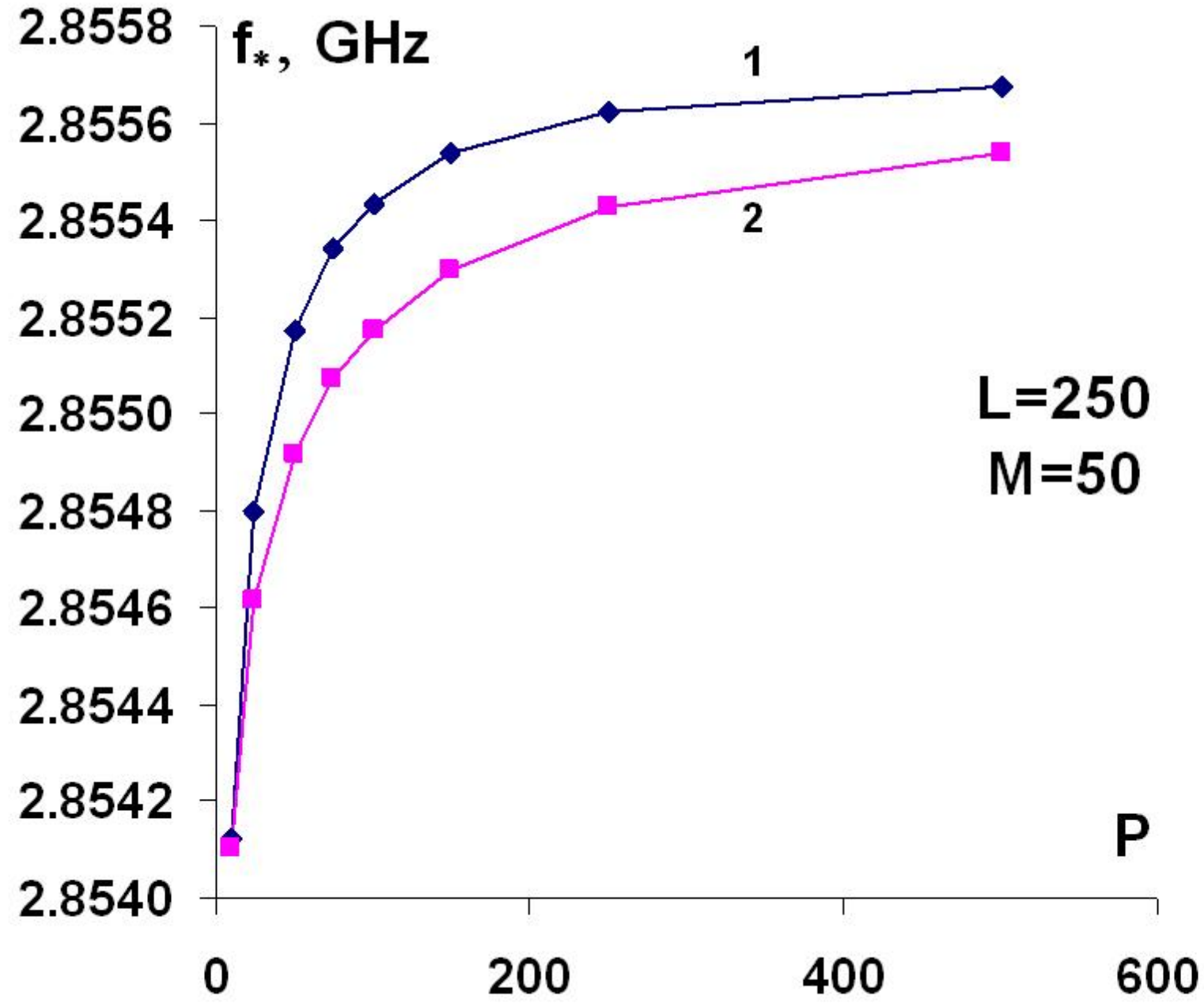

**Fig. 4 Dependence of frequency of $2\pi/3$ mode on number of homogeneous waveguides that represent the real opening in the disk ($2P+1$) for homogeneous DLW with $d=$3.099 cm, $t=$0.4 cm, $b=$4.134 cm, $a=$1.2 cm**

In Fig. 3 and Fig. 4 calculation results of frequency are presented with taking into account a coupling of nine cavities ($N=4$) for homogeneous DLW with $d=3.099$ cm, $t=0.4$ cm, $b=4.134$ cm, $a=1.2$ cm, $R=t/2=0.2$ cm (radius of rounding), SUPERFISH frequency of $2\pi/3$ mode is $f=$ 2855.06 MHz. We can see that solutions of the truncated set of equations converge. Two curves correspond to different laws of changing of lengths $l_p$ of homogeneous waveguides inside the opening. The first curve was obtained under the law $l_p = l_{p-1} + \Delta l$ ($l_0 = R/(10P)$) and the second one under the law $l_p = const$ ($\sum l_p = R = t/2$). Radii $a^{(p)}$ of homogeneous waveguides inside the opening were chosen from such condition

$$\pi\left(a^{(p)2} - a^2\right) l_p = V_p = \int_{z_p}^{z_{p+1}} dz \int_{R_0}^{a+R-\sqrt{R^2-(z-R)^2}} r dr d\varphi \qquad (1.76)$$

**Table 1**

| | $a=1.2$ cm, $b=4.134$ cm, $d=3.099$ cm, $t=2R=0.4$ cm |
|---|---|
| $\alpha_{1,0}^{(0,0)}$ | 0.04706432 |
| $\alpha_{1,0}^{(0,1)}$ | 0.01091948 |
| $\alpha_{1,0}^{(0,2)}$ | -1.9937E-05 |
| $\alpha_{1,0}^{(0,3)}$ | 3.7465E-08 |
| $\alpha_{1,0}^{(0,4)}$ | -7.0364E-11 |

Taking into account a coupling of nine cavities ($N=4$) for the homogeneous DLW with considered geometry[4] is excessive. Indeed, from Table 1 it follows that coefficient $\alpha_{1,0}^{(0,4)}$ is less than the needed accuracy. So, in our calculations we shall use $N=3$ approach for the DLWs with geometry close to considered one above.

[4] Taking into account a coupling of nine cavities ($N=4$) can be necessary for a DLW with greater coupling (small d or large a).

In paper [9] we presented some results that clarify some aspects of the tuning methods that are based on the field measurements and were used for tuning nonuniform DLWs [11,12,13,14,15,16,17,18,19] . These results concerned the possibility of using some parameters that characterize the detuning of cell without taking into account the parameters of neighboring cells. Using the above coupled cavity model we can make calculation of characteristics of the inhomogeneous disc-loaded waveguides that are used in real linacs with taking into account the rounding of the disk hole edges.

In paper [9] we proposed to characterize the sell detuning in the case of small coupling by such parameter

$$F_k^{(3)} = \frac{Z_k^{(m,3)}}{Z_k^{(c,3)}} - 1, \tag{1.77}$$

where[5]

$$Z_k^{(m,3)} = \frac{\alpha_{10}^{(c,k,k-1)} e_{010}^{(k-1,m)} + \alpha_{10}^{(c,k,k+1)} e_{010}^{(k+1,m)}}{e_{010}^{(k,m)}}, \tag{1.78}$$

$$Z_k^{(c,3)} = \frac{\alpha_{10}^{(c,k,k-1)} e_{010}^{(k-1,c)} + \alpha_{10}^{(c,k,k+1)} e_{010}^{(k+1,c)}}{e_{010}^{(k,c)}}, \tag{1.79}$$

$\alpha_{10}^{(c,k,k-1)}, \alpha_{10}^{(c,k,k+1)}$ - calculated coupling coefficients, $e_{010}^{(k,c)}$ - calculated complex amplitudes of $E_{010}$ mode, $e_{010}^{(k,m)} = \frac{1}{d_k} \int_{d_k} \bar{E}_k (r=0,z) dz$ - measured amplitudes of $E_{010}$ mode, $\bar{E}_k$ - measured longitudinal electric field of the $k$-cavity.

In the case of moderate coupling we can use such parameter

$$F_k^{(5)} = \frac{Z_k^{(m,5)}}{Z_k^{(c,5)}} - 1. \tag{1.80}$$

$$Z_k^{(m,5)} = \frac{\alpha_{10}^{(c,k,k-2)} e_{010}^{(k-2,m)} + \alpha_{10}^{(c,k,k-1)} e_{010}^{(k-1,m)} + \alpha_{10}^{(c,k,k+1)} e_{010}^{(k+1,m)} + \alpha_{10}^{(c,k,k+2)} e_{010}^{(k+2,m)}}{e_{010}^{(k,m)}}, \tag{1.81}$$

$$Z_k^{(5)} = \frac{\alpha_{10}^{(c,k,k-2)} e_{010}^{(k-2,c)} + \alpha_{10}^{(c,k,k-1)} e_{010}^{(k-1,c)} + \alpha_{10}^{(c,k,k+1)} e_{010}^{(k+1,c)} + \alpha_{10}^{(c,k,k+2)} e_{010}^{(k+2,c)}}{e_{010}^{(k,c)}}. \tag{1.82}$$

There also was proposed to use such parameter [11,12,13,14,15,16,17,18,19]

$$F_k^{(3,0)} = \frac{\bar{E}_{k-1}^{(m)} + \bar{E}_{k+1}^{(m)} - 2\cos\varphi \bar{E}_k^{(m)}}{\bar{E}_k^{(m)} 2\cos\varphi}, \tag{1.83}$$

where $\bar{E}_k^{(m)} = \bar{E}_k (r=0, z=d_k/2)$ measured electric field in the center of the $k$-cavity. We also introduced an additional parameter that can be obtained from (1.77) under assumption that $Z_k^{(c,3)} \approx 2\alpha_{10}^{(c,k,k-1)} \cos\varphi, \;\; \alpha_{10}^{(c,k,k-1)} \approx \alpha_{10}^{(c,k,k+1)}$ [9] (compare with (1.83))

$$F_k^{(3,00)} = \frac{e_{010}^{(k-1,m)} + e_{010}^{(k+1,m)} - 2\cos\varphi \, e_{010}^{(k,m)}}{2\cos\varphi \, e_{010}^{(k,m)}}. \tag{1.84}$$

[5] c- calculated values, m – measured ones

To study quality of introduced parameters we calculated these parameters for a homogeneous DLW[6] with $\beta_{ph} \approx 1$ ($d_k^{(0)} = 3.099$ cm, $t_k^{(0)} = 0.4$ cm, $b_k^{(0)} = 4.1334$ cm, $a_k^{(0)} = 1.2056$ cm, $R_k^{(0)} = t/2 = 0.2$ cm, SUPERFISH frequency of $2\pi/3$ mode $f =$ 2856.04 MHz, frequency of $2\pi/3$ mode under calculation with $P = 50$, $L = 250$, $M = 50$ - $f =$ 2856.38 MHz) that has a local "defect". We used the developed coupled cavity model with $N =$ 3 for calculation the necessary coefficients.

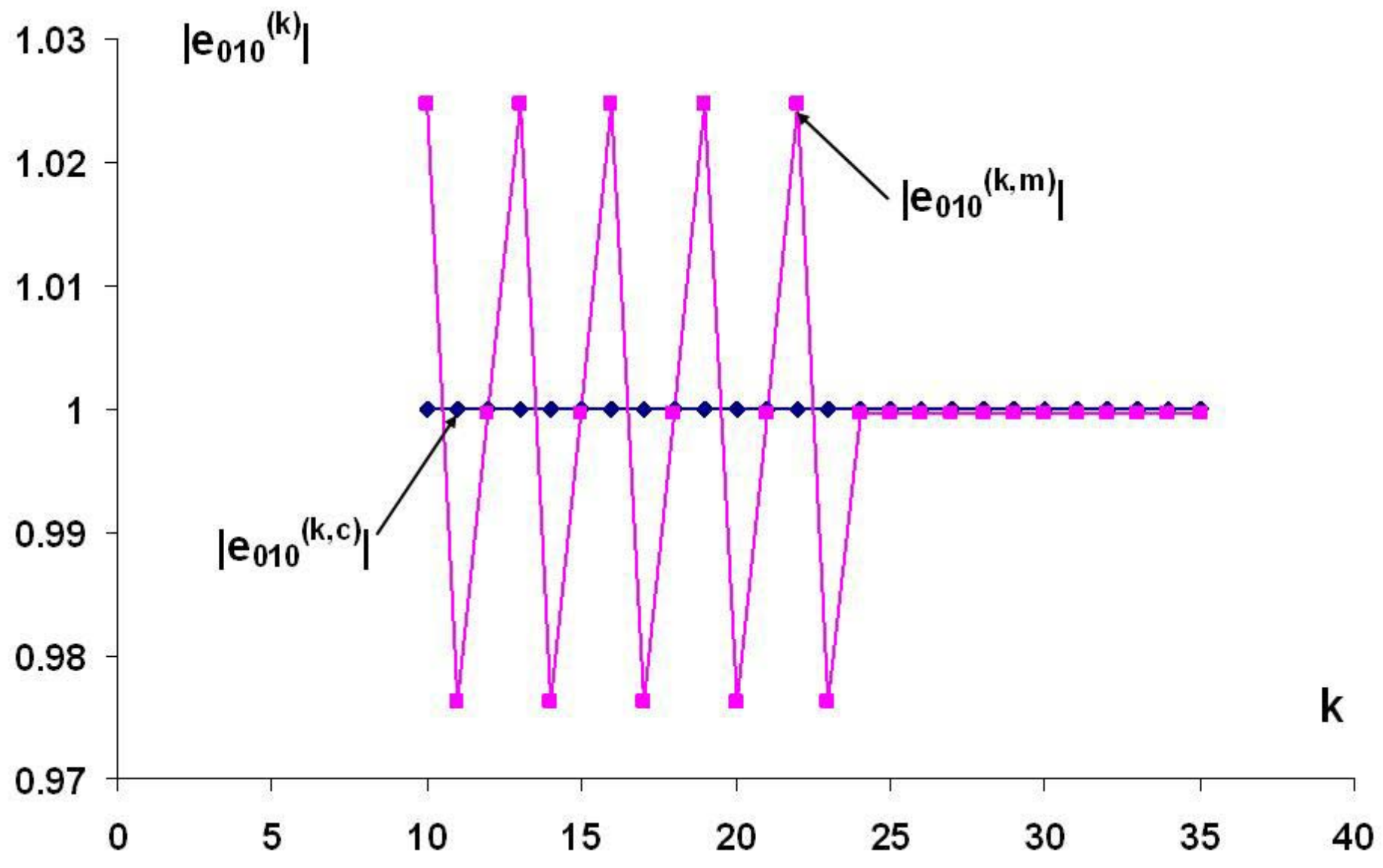

**Fig. 5 Modules of amplitudes of $E_{010}$ mode in the case of the "b-defect": $b_{24} = b_{24}^{(0)} - 10\mu m$**

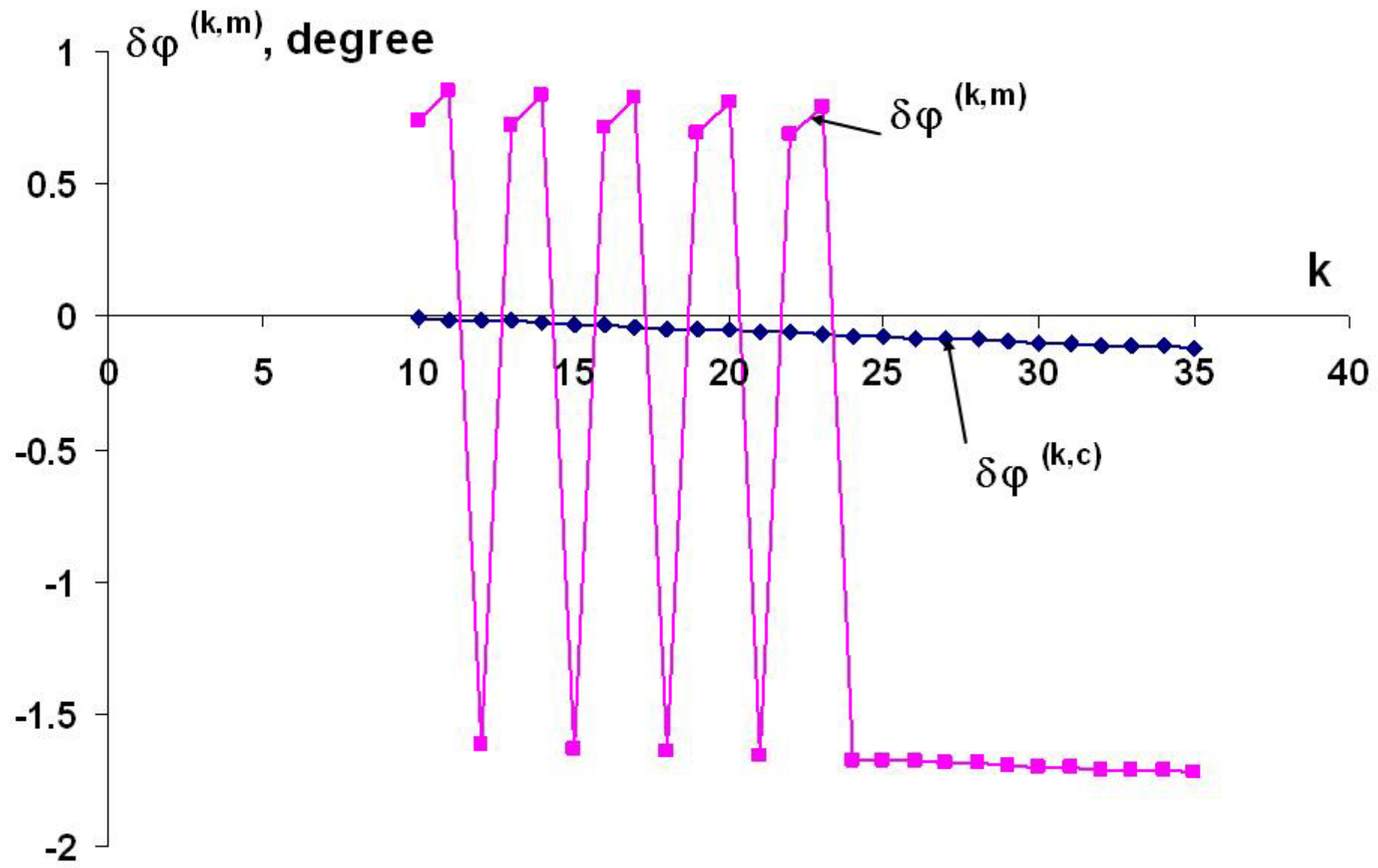

**Fig. 6 Phase shifts ($\varphi_k - 120^\circ$) in the case of the "b-defect": $b_{24} = b_{24}^{(0)} - 10\mu m$**

Calculations were carried out for 25 cavities with assumption that at frequency $f =$ 2856.38 MHz there is a full matching at the ends [8]. Amplitudes and phase shifts ($\varphi_k - 120^\circ$) for the DLW without defects are presented in Fig. 5 and Fig. 6 (coefficients $\left|e_{010}^{(k,c)}\right|$ and $\delta\varphi^{(k,c)}$). Reflection coefficient equals 8E-5; real parts of $F_k^{(3,00)}$ and $F_k^{(3,0)}$ equal:

[6] The last cell of an accelerating structure for industrial linac ( M.I. Ayzatskiy, A.N. Dovbnya, V.F. Ziglo et al. Accelerating system for an industrial linac. PAST, 2012, N4, pp.24-28)

$\text{Real}(F_k^{(3,00)}) = \text{Real}(F_k^{(3,0)}) =$ -1.39E-4; modules of real parts of $F_k^{(3,\text{m})}$ and $F_k^{(5,\text{m})}$ are less than E-14; imaginary parts are very small for all coefficients (modules<E-14).

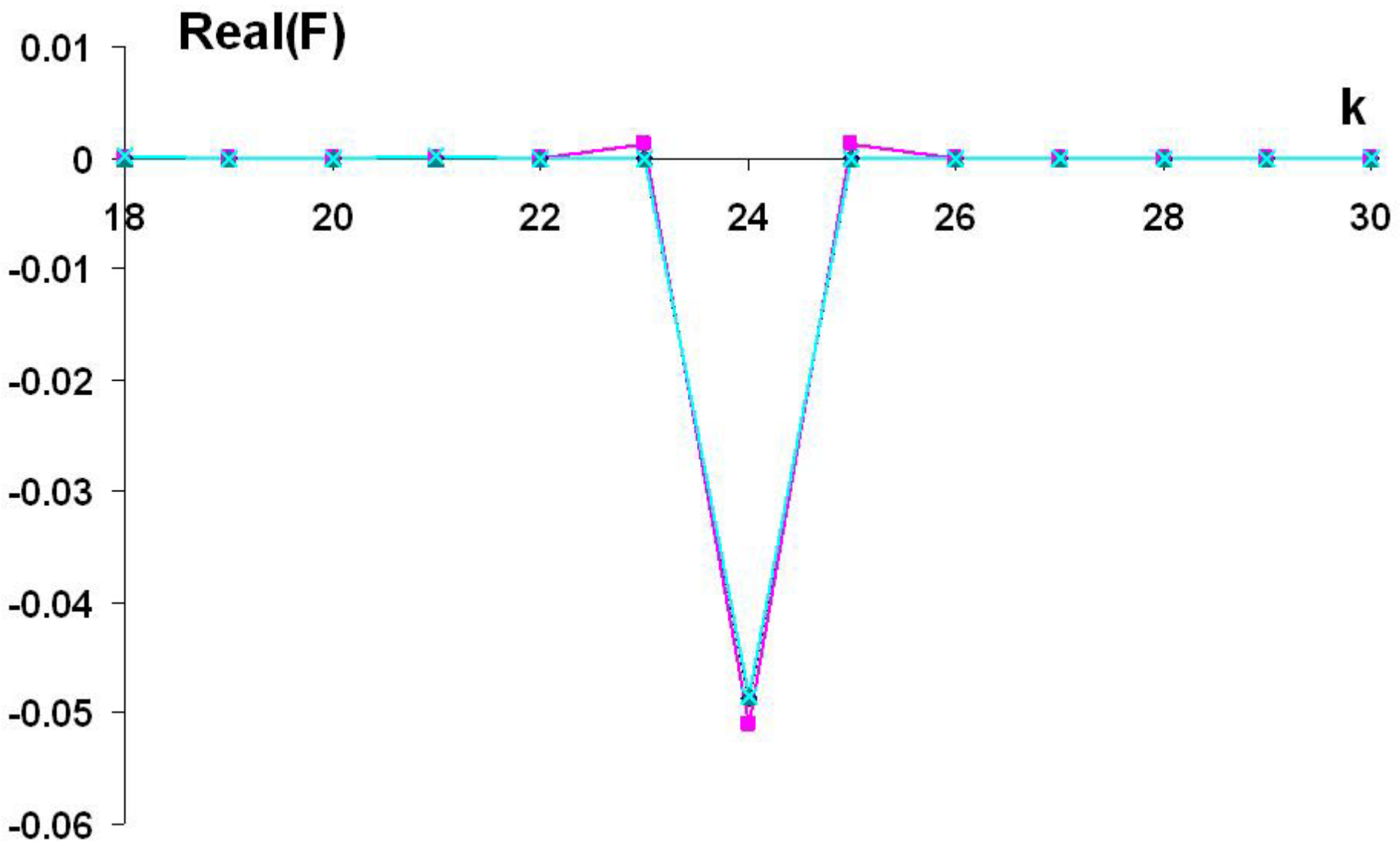

**Fig. 7 Real parts of parameters $F_k^{(3,00)}, F_k^{(3,0)}, F_k^{(3)}, F_k^{(5)}$ in the case of the "b-defect":** $b_{24} = b_{24}^{(0)} - 10\mu m$

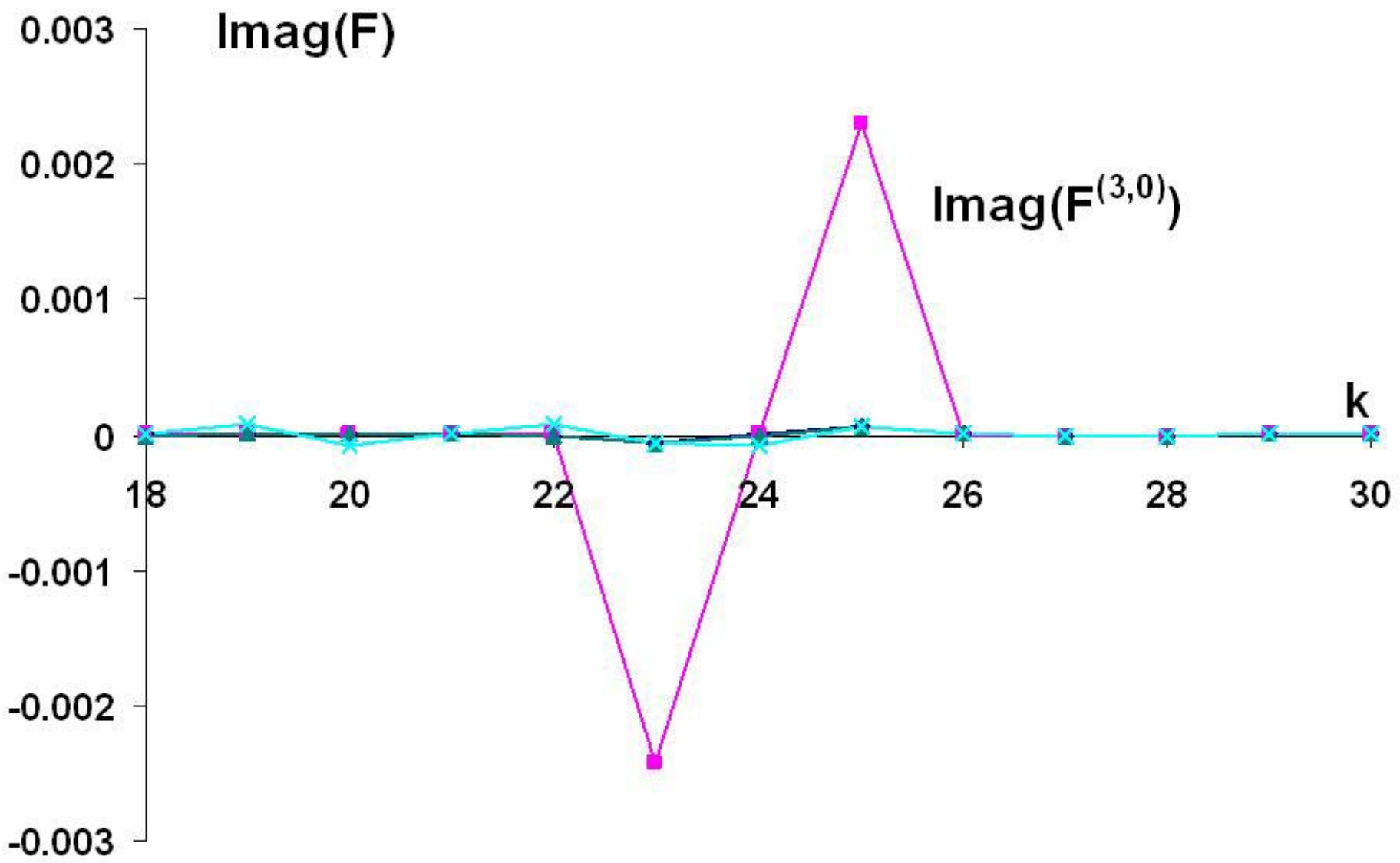

**Fig. 8 Imaginary parts of parameters $F_k^{(3,00)}, F_k^{(3,0)}, F_k^{(3)}, F_k^{(5)}$ in the case of the "b-defect":** $b_{24} = b_{24}^{(0)} - 10\mu m$

First of all, we made calculation for the case when the DLW has the "b-defect" – deviation of cell radius: $b_{24} = b_{24}^{(0)} - 10\mu m$. We remind that the tolerances of a SLAC 10 feet section were: $\Delta 2b = \pm$ 1.3 µm, $\Delta 2a = +$ 5 µm, $\Delta d = \pm 25$ µm, $\Delta t = \pm$ 5 µm, $\Delta R = \pm$ 13 µm [20]. Amplitudes and phase shifts ($\varphi_k - 120^\circ$ ) are presented in Fig. 5 and Fig. 6 . The reflection coefficient in this case equals 2.8E-2. Real and imaginary parts of all four parameters $F_k^{(3,00)}, F_k^{(3,0)}, F_k^{(3)}, F_k^{(5)}$ are presented in Fig. 7 and Fig. 8. Real parts of all parameters have close values. Imaginary parts of three parameters $F_k^{(3,00)}, F_k^{(3)}, F_k^{(5)}$ are small, while the imaginary parts of parameters[7] $F_{22}^{(3,0)}$ and $F_{25}^{(3,0)}$ differ from the imaginary parts of other parameters. But when we will try to eliminate this defect (to make a real part of this parameters smaller), the imaginary part of parameters $F_{22}^{(3,0)}$ and $F_{25}^{(3,0)}$ will tend to zero, too.

In the case of "a-defect" situation becomes more complicated as this "defect" can not be removed under section tuning by changing cell radii. Modules of $E_{010}$ amplitudes and phase shifts[8] ($\varphi_k - 120^\circ$) for the case $a_{24} = a_{24}^{(0)} - 10\mu m$ are presented in Fig. 9 and Fig. 10. The reflection coefficient in this case equals 6E-3.

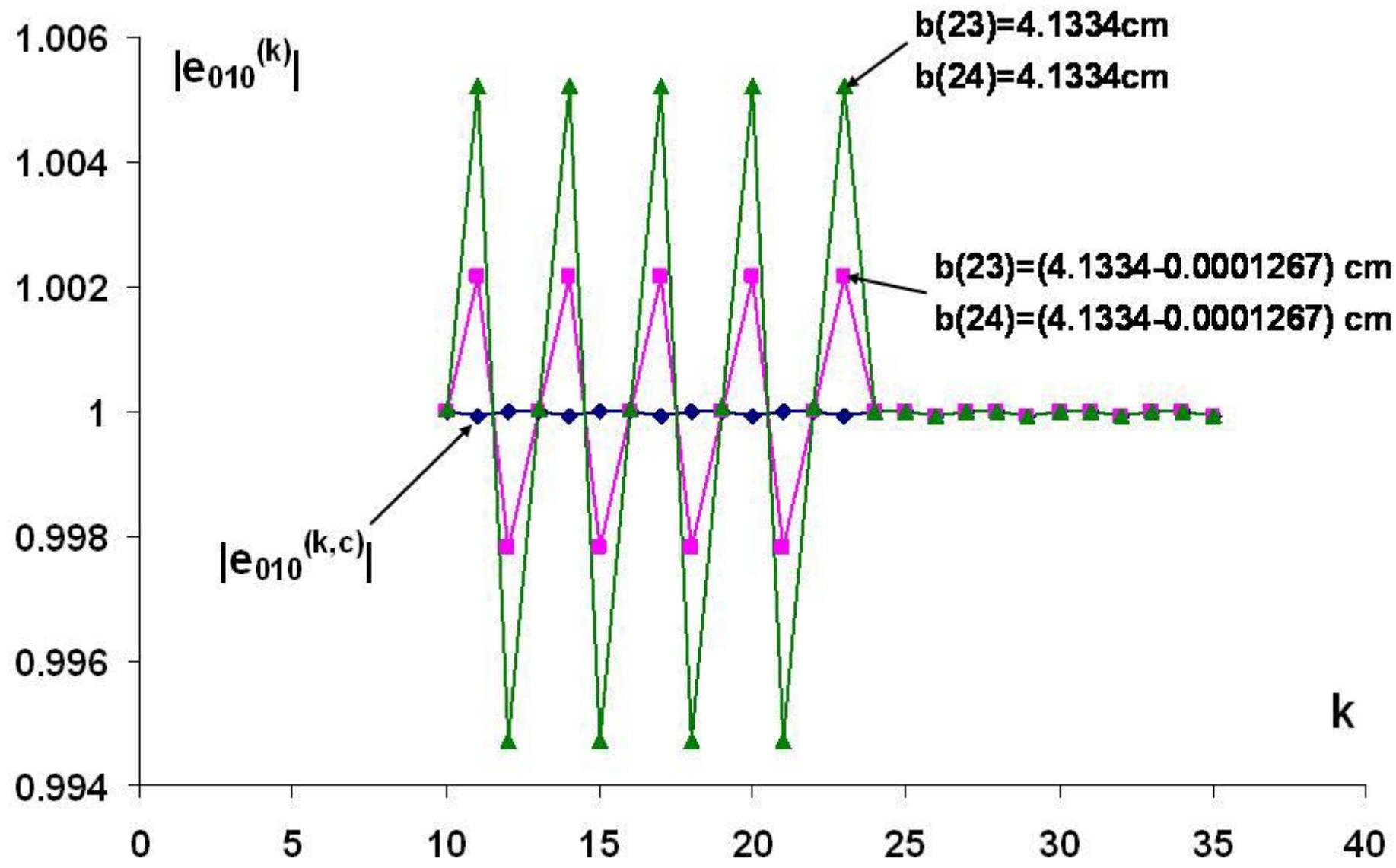

**Fig. 9 Modules of amplitudes of $E_{010}$ mode in the case of the "a-defect":** $a_{24} = a_{24}^{(0)} - 10\mu m$

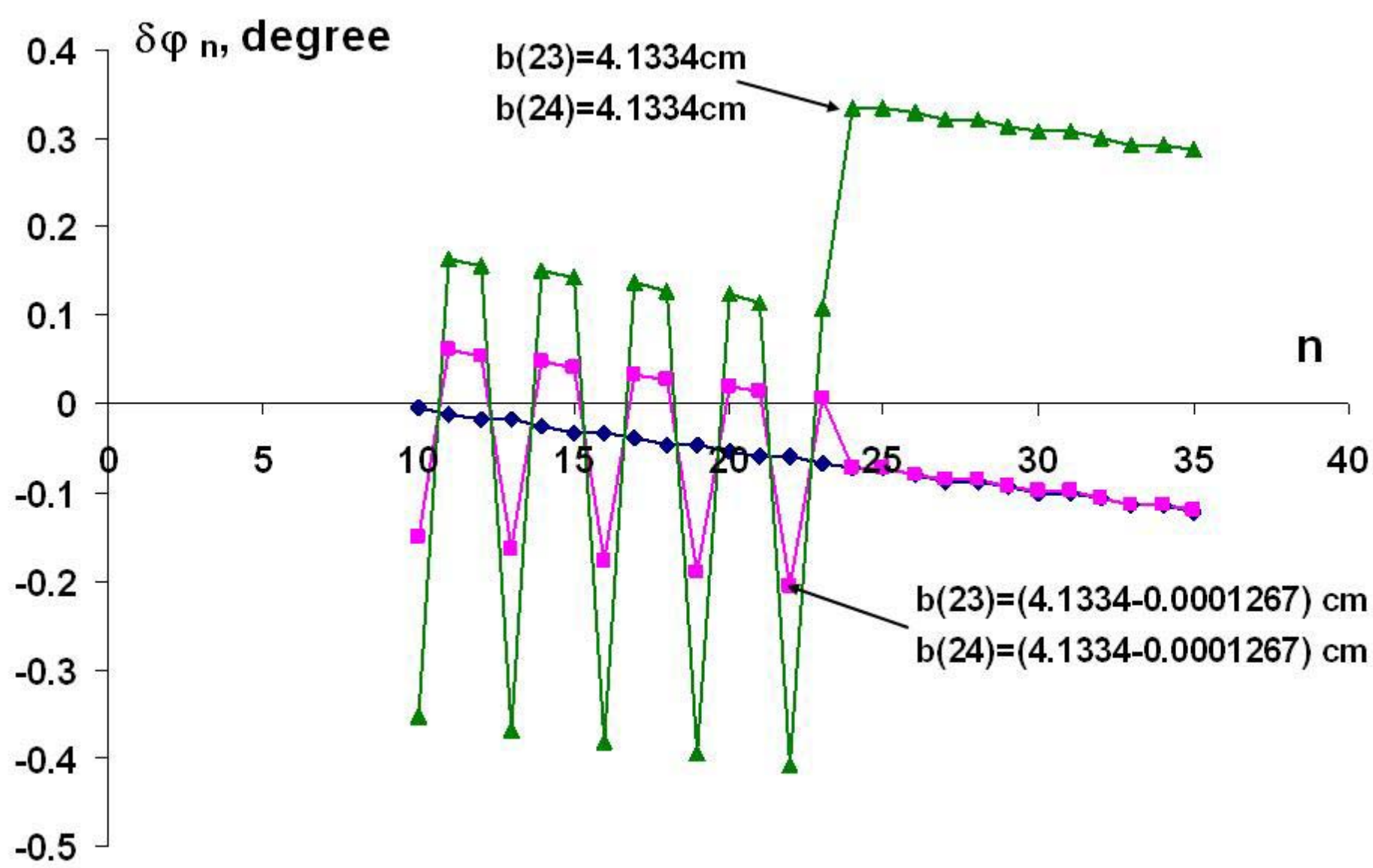

**Fig. 10 Phase shifts ($\varphi_k - 120^\circ$) in the case of the "a-defect":** $a_{24} = a_{24}^{(0)} - 10\mu m$

[7]It is defined through measured electric field in the center of a cavity, see (1.83)

[8] Phase shifts for $E_{010}$ amplitudes and electric fields in the center of cells have nearly the same values

We present in Fig. 11 values of module electric field in the center of cavities for explaining why the parameter $F_k^{(3,0)}$ always differs from other parameters $F_k^{(3,00)}, F_k^{(3)}, F_k^{(5)}$. Comparing results in Fig. 9 and Fig. 11, we can see that fields in the centers of disturbed cells (cells N23 and 24) differ from fields in the corresponding cells that lay on the right and the left (Fig. 11) in contrast with the amplitudes of $E_{010}$ mode (Fig. 9).

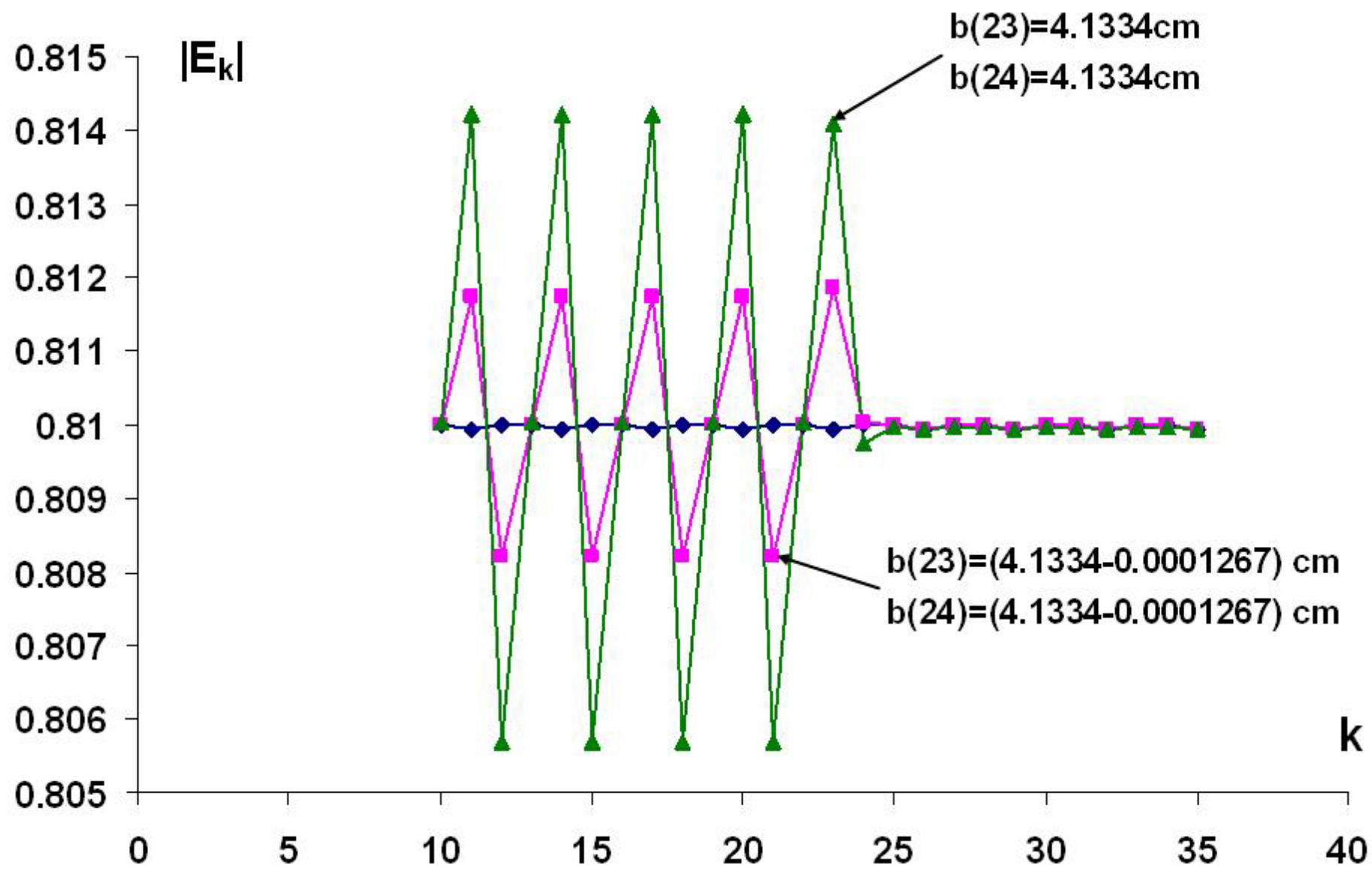

**Fig. 11 Modules of fields in the centers of cells in the case of the "a-defect":** $a_{24} = a_{24}^{(0)} - 10\mu m$

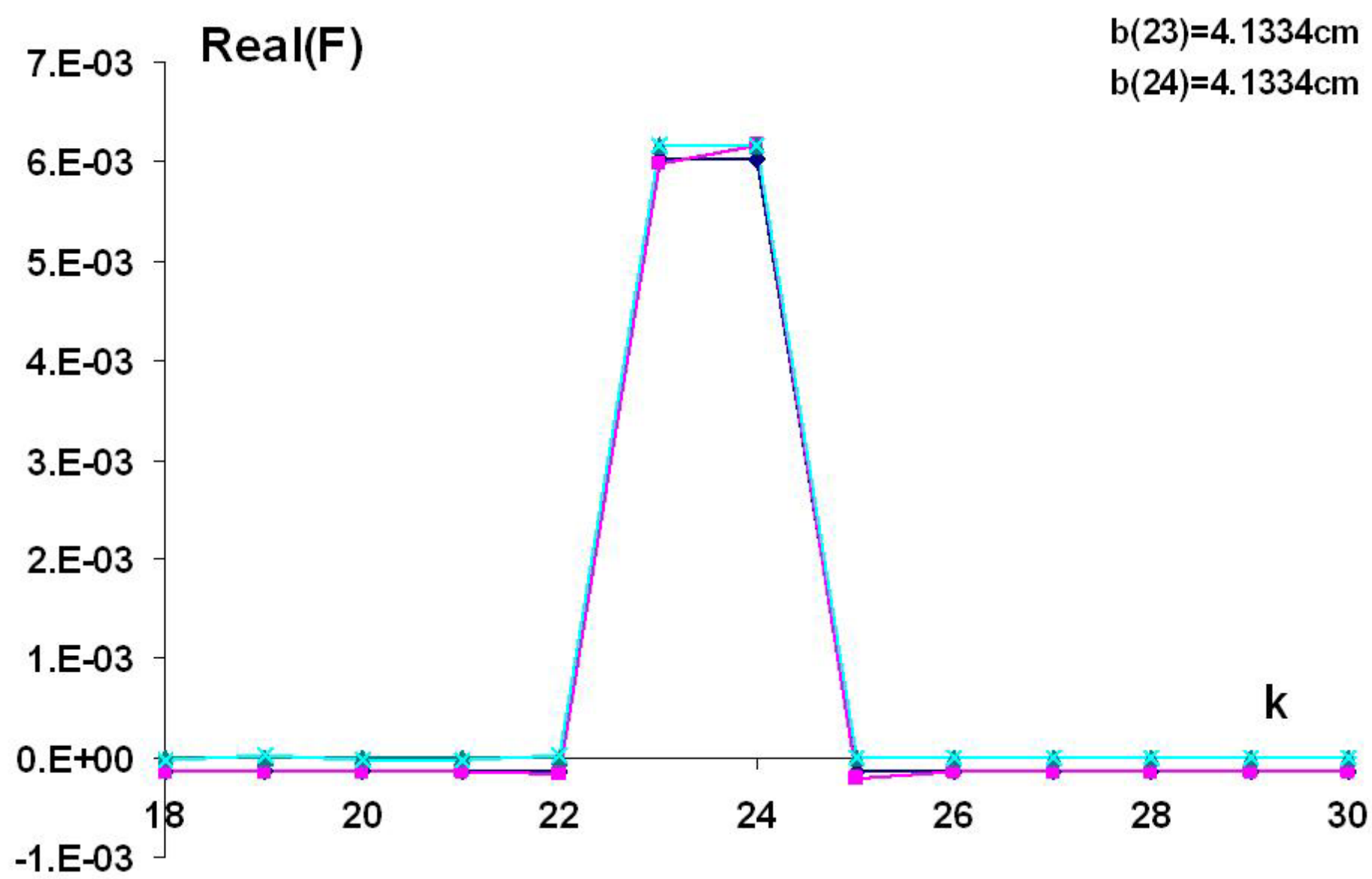

**Fig. 12 Real parts of parameters** $F_k^{(3,00)}, F_k^{(3,0)}, F_k^{(3)}, F_k^{(5)}$ **in the case of the "a-defect":** $a_{24} = a_{24}^{(0)} - 10\mu m$

Real and imaginary parts of all four parameters $F_k^{(3,00)}, F_k^{(3,0)}, F_k^{(3)}, F_k^{(5)}$ are presented in Fig. 12 and Fig. 13. Real and imaginary parts of all parameters have close values. It has to be noted that for this kind of defect imaginary parts of all parameters are not close to zero. Moreover, imaginary parts of all parameters practically do not change under the changing of cell radii (compare Fig. 13 and Fig. 15). When we will try to make a real part of these parameters smaller by changing $b_{23}$ and $b_{24}$ (see Fig. 14), it leads to decreasing the reflection coefficient up to 2.5E-3. But the zero value of the reflection coefficient can be obtained when the real parts of parameters $F_k^{(3,00)}, F_k^{(3,0)}, F_k^{(3)}, F_k^{(5)}$ are not close to zero. So, we can conclude that for this kind of defect making the real parts of parameters $F_k^{(3,00)}, F_k^{(3,0)}, F_k^{(3)}, F_k^{(5)}$ close to zero we decrease reflections but do not fully compensate them.

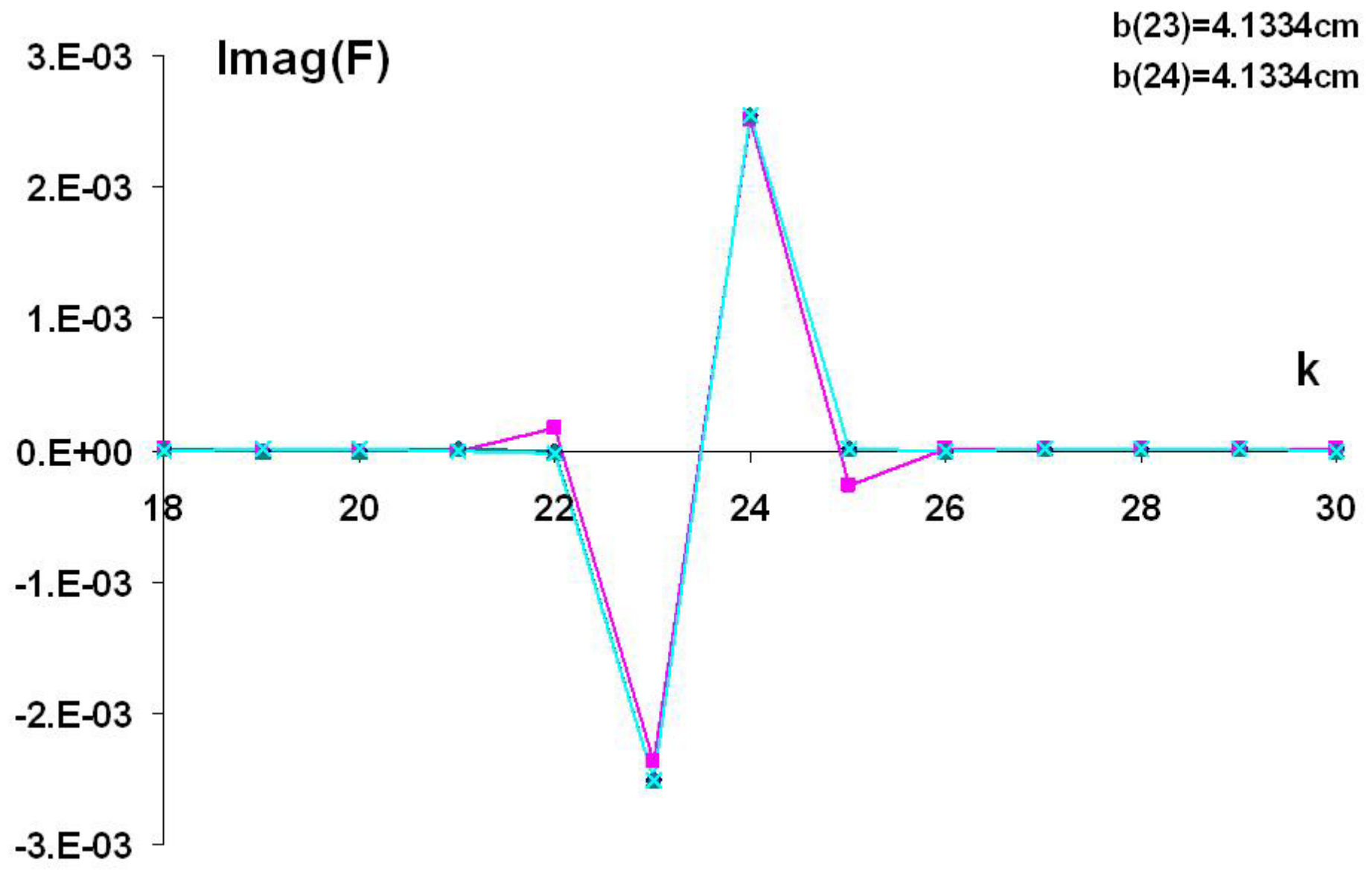

**Fig. 13 Imaginary parts of parameters $F_k^{(3,00)}, F_k^{(3,0)}, F_k^{(3)}, F_k^{(5)}$ in the case of the "a-defect":** $a_{24} = a_{24}^{(0)} - 10\mu m$

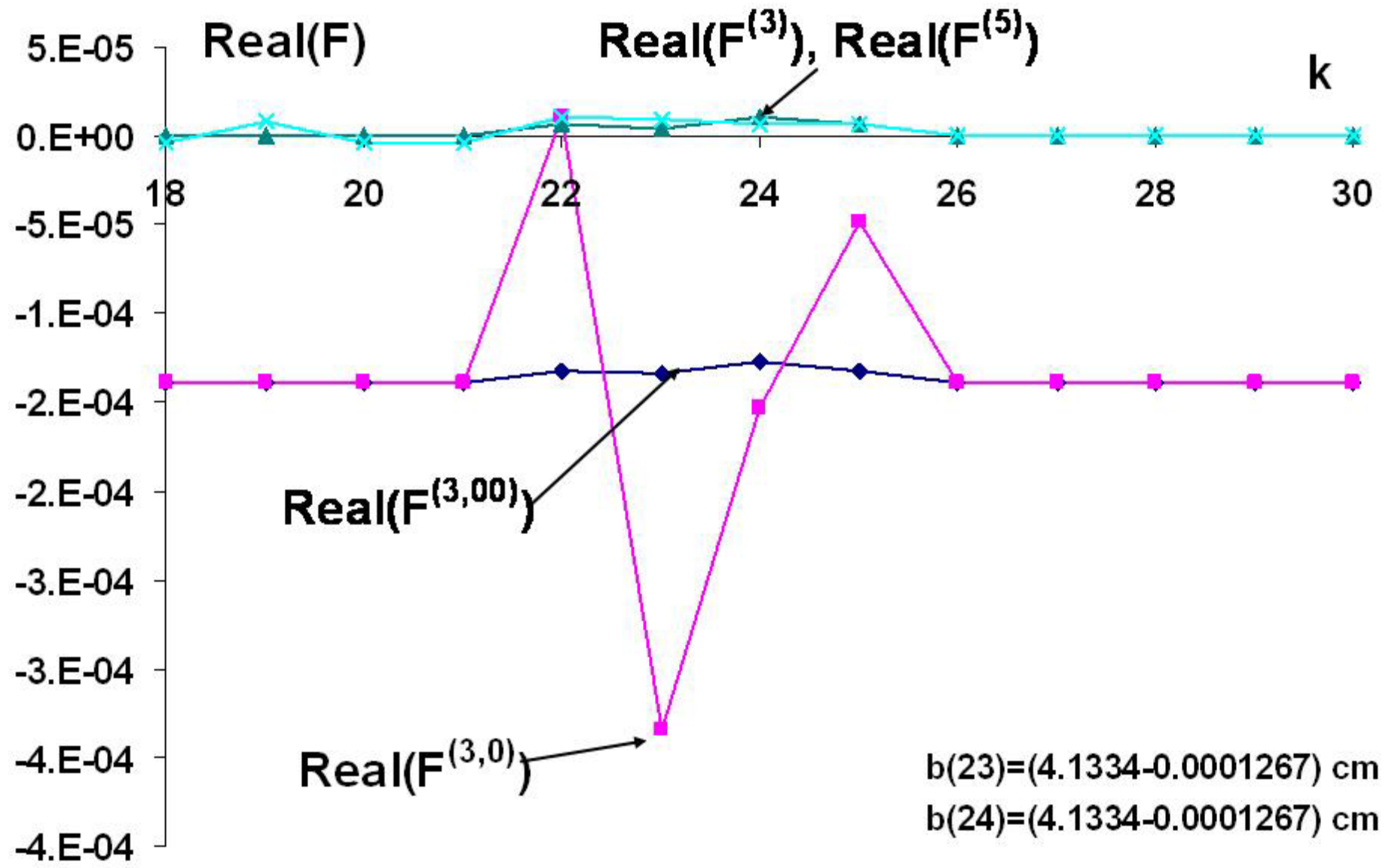

**Fig. 14 Real parts of parameters $F_k^{(3,00)}, F_k^{(3,0)}, F_k^{(3)}, F_k^{(5)}$ in the case of the "a-defect":** $a_{24} = a_{24}^{(0)} - 10\mu m$ **under changing** $b_{23}$ **and** $b_{24}$

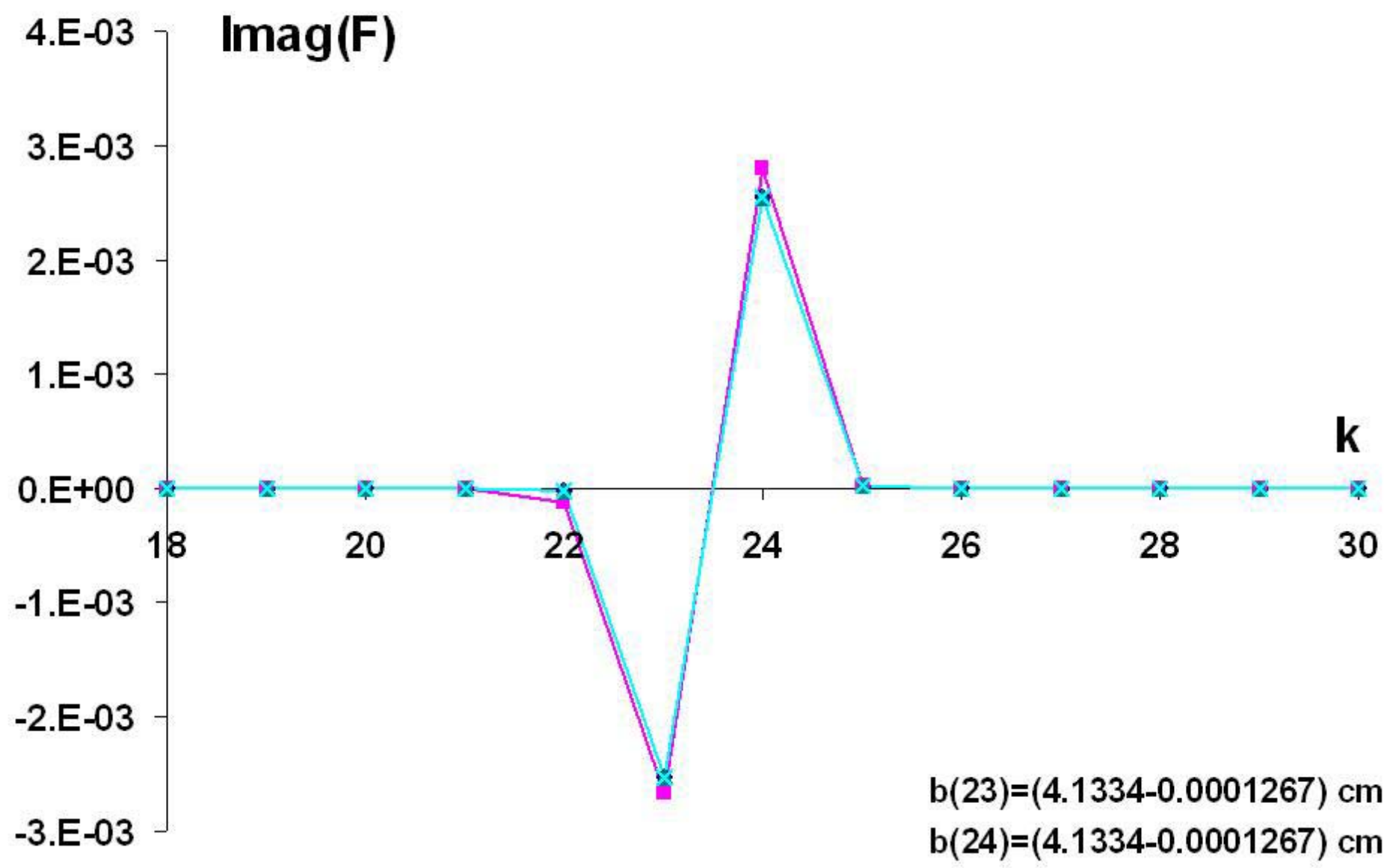

**Fig. 15 Imaginary parts of parameters $F_k^{(3,00)}, F_k^{(3,0)}, F_k^{(3)}, F_k^{(5)}$ in the case of the "a-defect": $\tilde{a}_{24} = a_{24} - 10\mu m$ under changing $b_{23}$ and $b_{24}$**

We also conducted simulation of cells tuning for other types of defect: $t, d, R1, R2$. Under $R1$ and $R2$ defects we understand the deviation in radii of rounding of hole edges. For all types of defects imaginary parts of all introduced parameters are not close to zero and practically do not change under the changing of cell radii. Decreasing the real parts of parameters $F_k^{(3,00)}, F_k^{(3,0)}, F_k^{(3)}, F_k^{(5)}$, we decrease reflections but do not fully compensate them.

As we have already noted [8,9], proposed model can be used for tuning inhomogeneous DLW. In some cases one can use simple coefficients $F_k^{(3,00)}, F_k^{(3,0)}$, which can be obtained without numerical simulation. But in general case, it is needed to conduct full numerical simulation of specific DLW and obtain all necessary coupling coefficients and amplitudes. After that one can start the tuning process by using the $F_k^{(3)}$ coefficient (or even the $F_k^{(5)}$ coefficient).

We shall illustrate this process by such example. Let's suppose that we want to develop a homogeneous DLW with phase shift. This shift can be realized by changing the radius of some disk and choosing the appropriate values of radii of adjacent cells. For the homogeneous DLW that has dimensions equal to those that was considered above we made simulation, results of which are presented in Fig. 16 and Fig. 17. The 24th hole has radius $a_{24}^{(0)} = (1.2056 - 0.01)cm$, The reflection coefficient equals 1.5E-4 for such radii of adjacent cells: $b_{24}^{(0)} = b_{23}^{(0)} = 4.1334 - 0.002225$. Using the calculated values of coupling coefficients and amplitudes we calculated the values of coefficients $F_k^{(3,00)}, F_k^{(3,0)}, F_k^{(3)}, F_k^{(5)}$ which are presented in Fig. 18 and Fig. 19.

We can see that only the coefficients $F_k^{(3)}, F_k^{(5)}$ have the zero values (both real and imaginary parts) in the tuned state. The coefficients $F_k^{(3,0)}$ has slightly different values in the cells (N23 and 24) that locate near the 24th hole. It is take place as a distribution of electric field in the centers of cells is nonuniform (see Fig. 16).

The main condition for possibility of using coefficients as criterion of cell tuning is independence (or very small dependence) of coefficient value for k-cell on the coefficient values for other cells ($\neq k$). In Fig. 20 results of calculation of $F_k^{(3)}$ coefficients are presented in the case of detuning of one cell by introducing "b-defect". Calculations were made for detuning the cell N22 (that locate near the inhomogeneity at the left), the cell N25 (that locate near the inhomogeneity at the right), and the cell N28. We can see that b-deviation in other cell practically do not influence on the values of $F_k^{(3)}$ coefficient in the inhomogeneity (cells with number $k=23,24$) The value of impact of other cells are illustrated by Table 2 and Fig. 21 where the imaginary parts of parameters $F_k^{(3)}$ in the case of "b-defect" are presented. We can see that such influence can be estimated on the level of E-4. So, we can conclude that $F_k^{(3)}$ coefficients for considered DLW parameters can be used in tuning process (it is also true for $F_k^{(5)}$ coefficients).

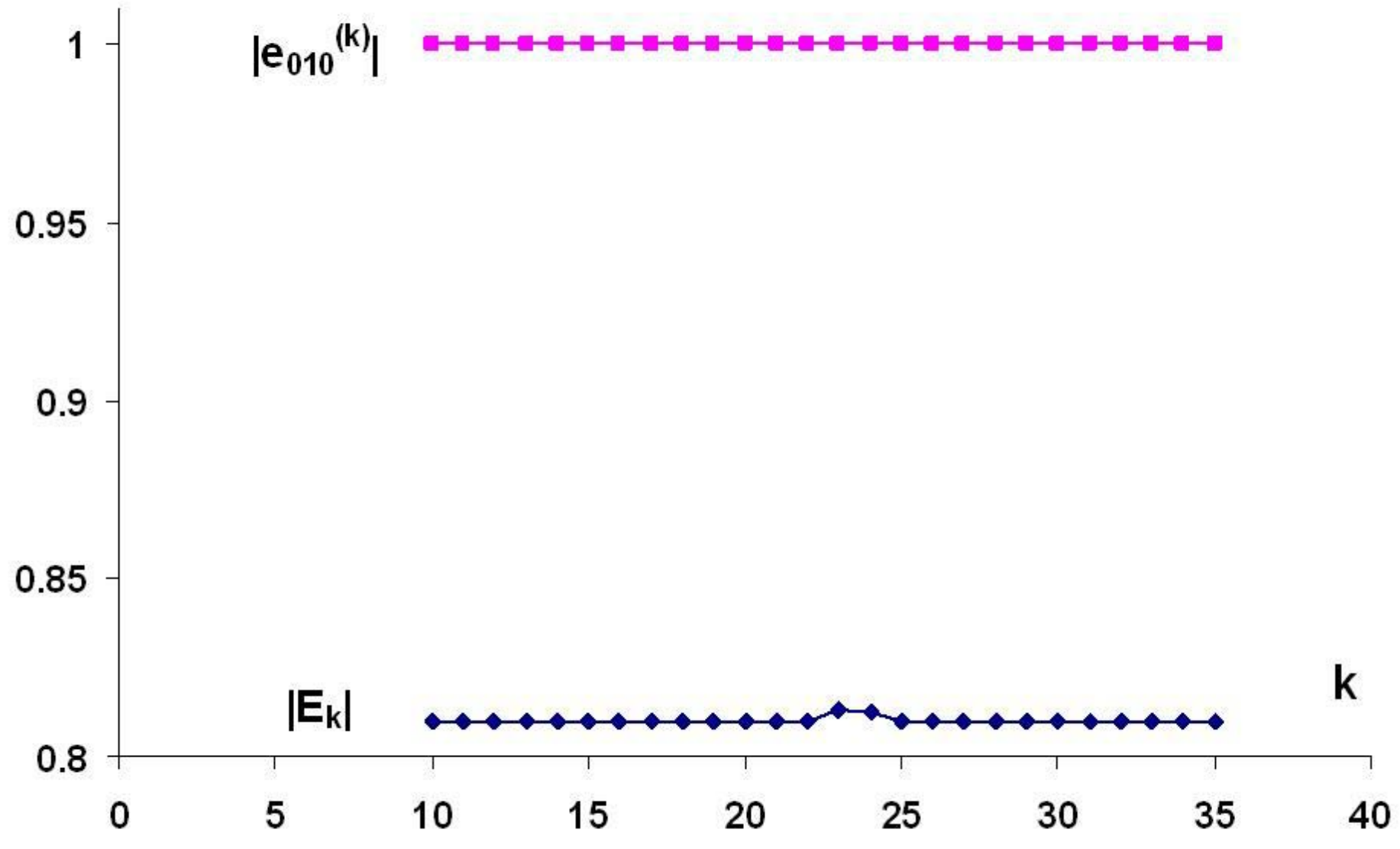

**Fig. 16 Modules amplitudes of $E_{010}$ mode and fields in the centers of cells**

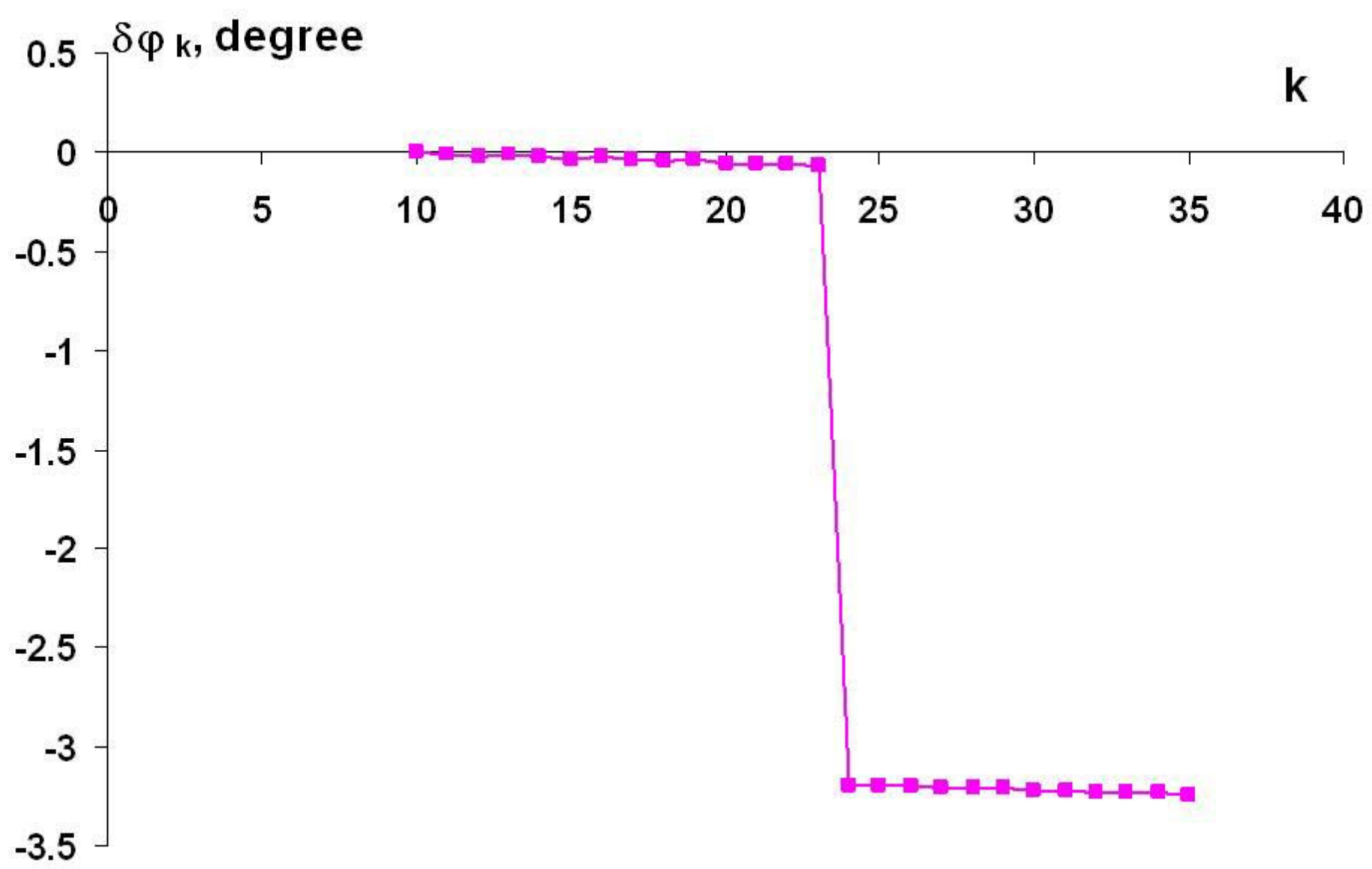

**Fig. 17 Phase shifts ($\varphi_k - 120^\circ$) in the centers of cells**

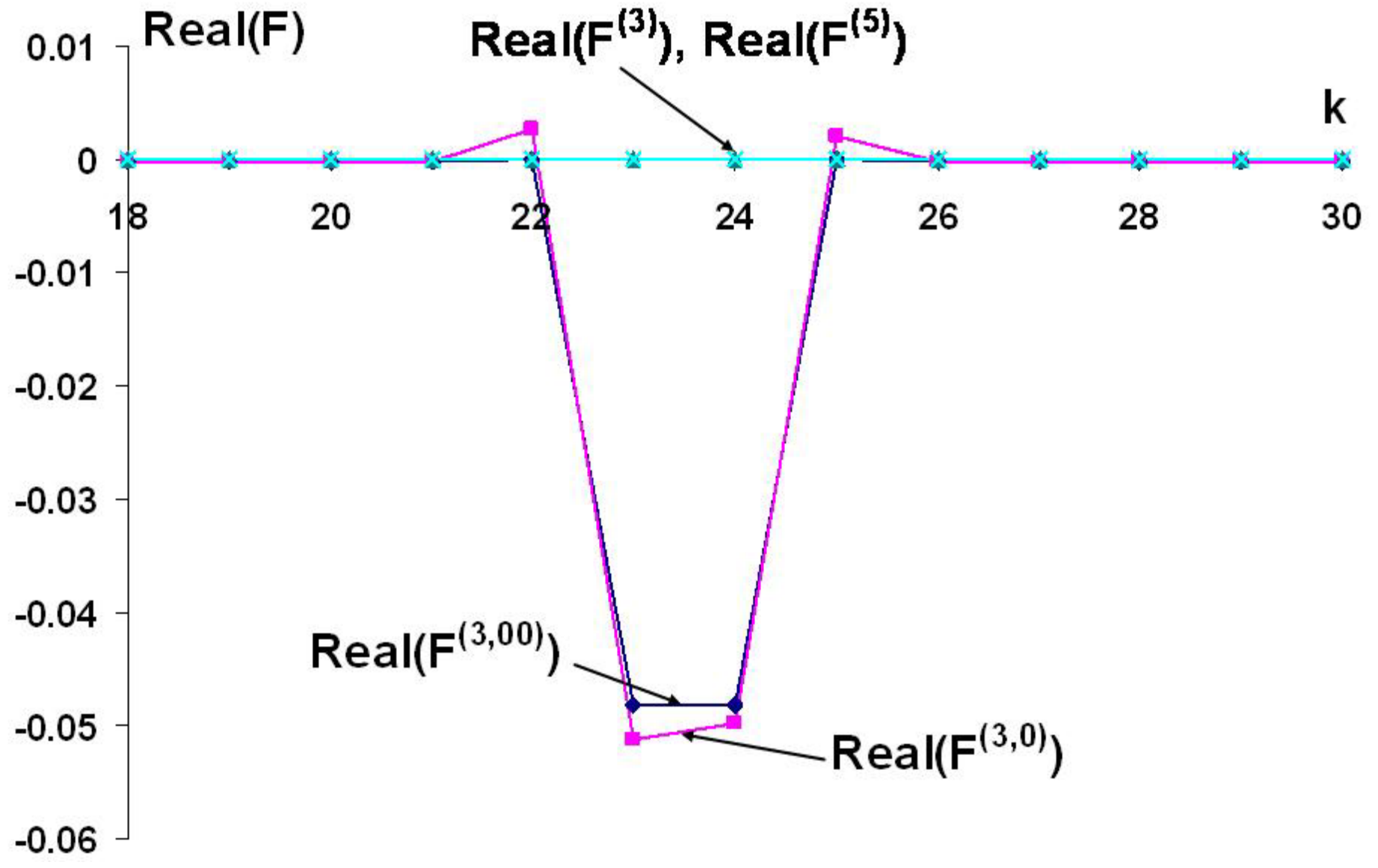

**Fig. 18 Real parts of parameters** $F_k^{(3,00)}, F_k^{(3,0)}, F_k^{(3)}, F_k^{(5)}$

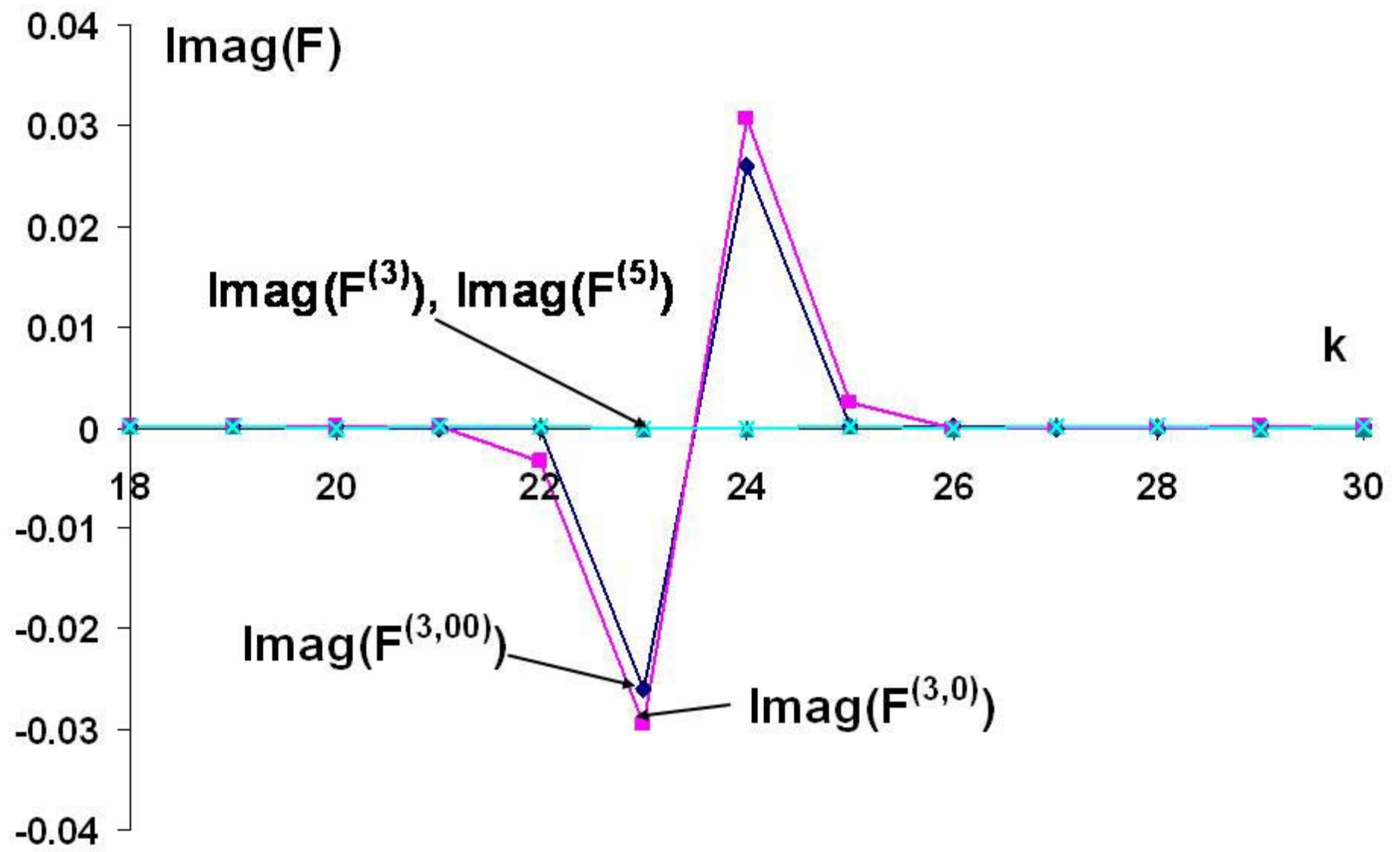

**Fig. 19 Imaginary parts of parameters** $F_k^{(3,00)}, F_k^{(3,0)}, F_k^{(3)}, F_k^{(5)}$

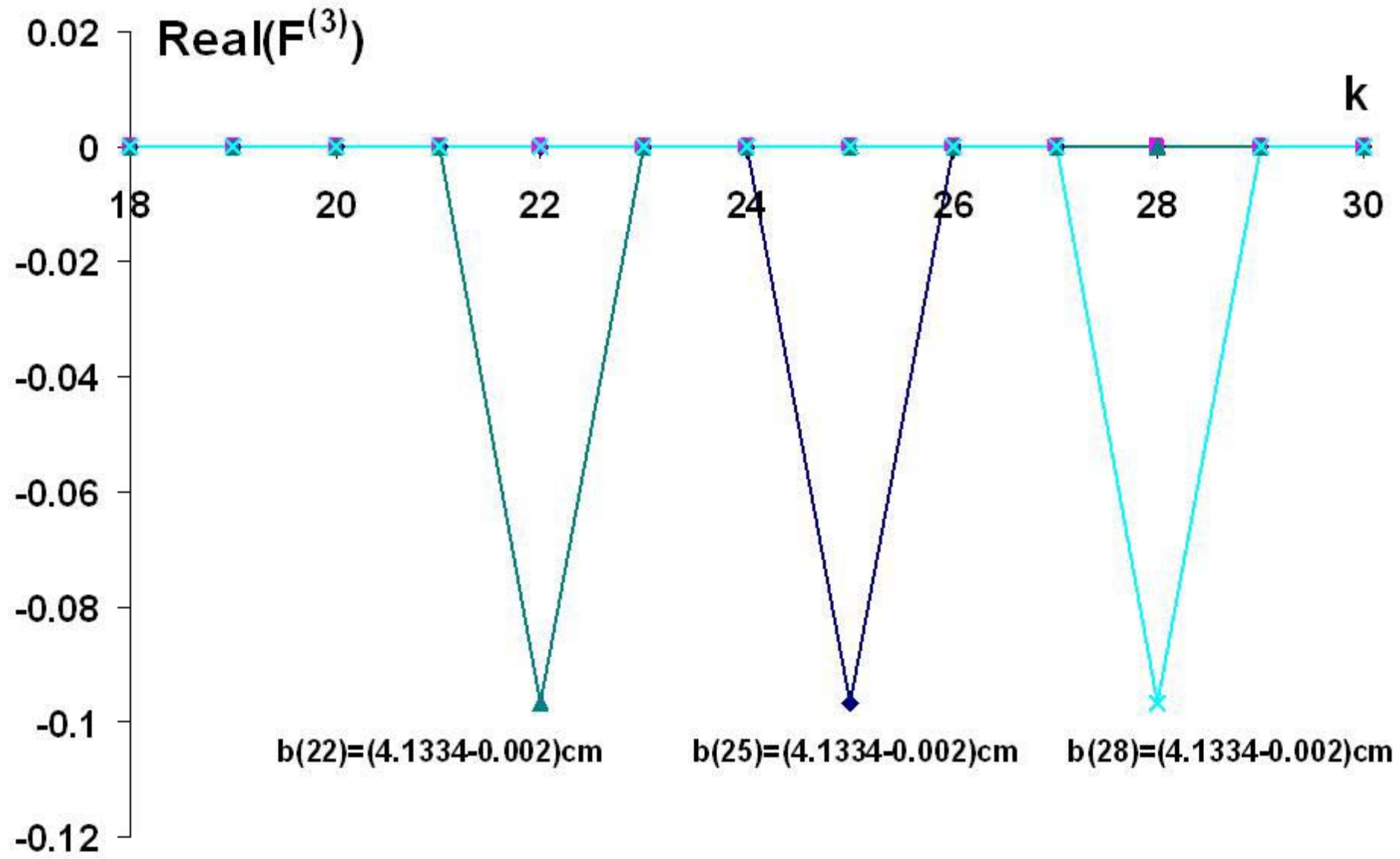

**Fig. 20 Real parts of parameter $F_k^{(3)}$ in the case of "b-deffect" in one cell that locates near the inhomogeneity**

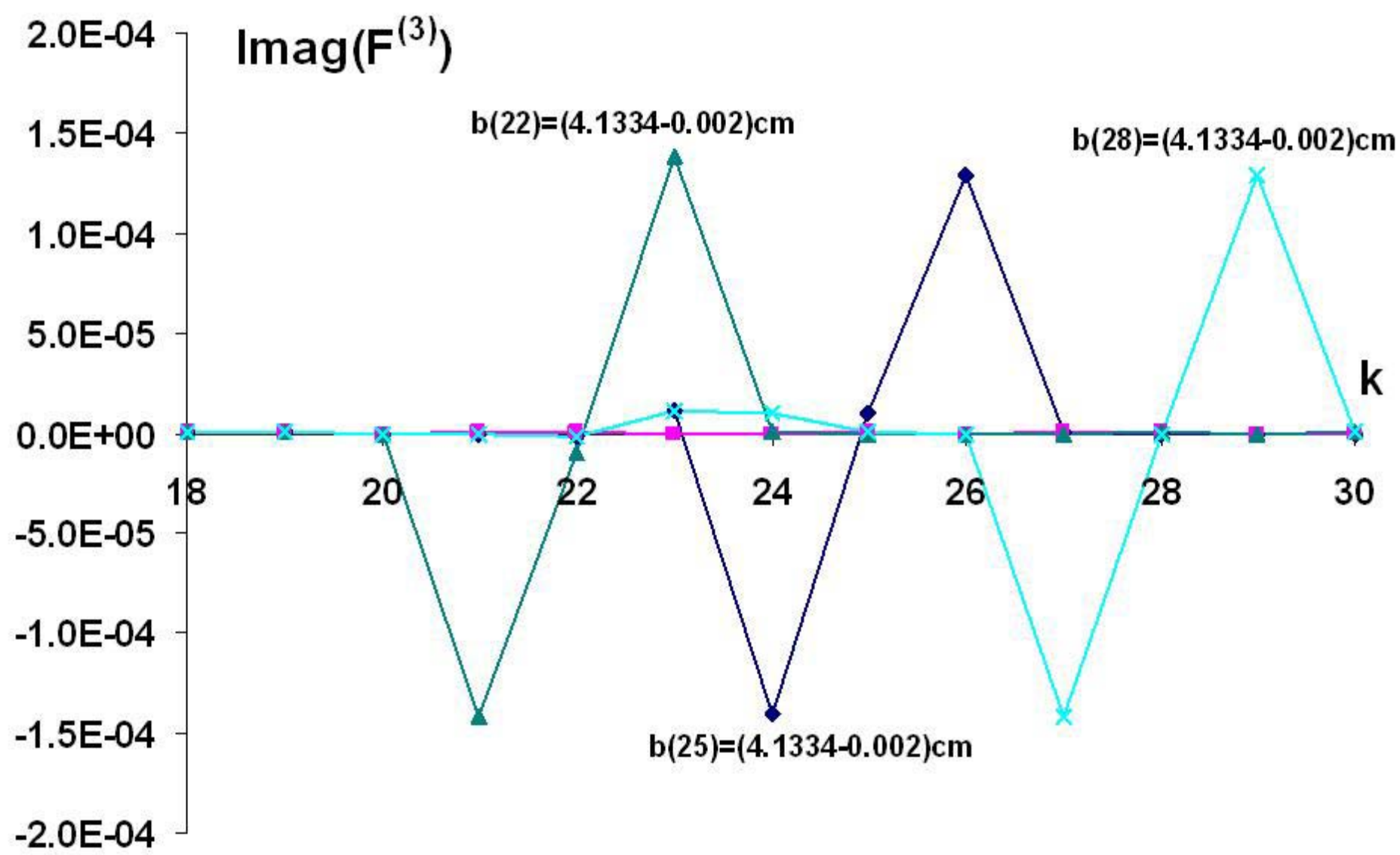

**Fig. 21 Imaginary parts of parameter $F_k^{(3)}$ in the case of "b-deffect" in one cell that locates near the inhomogeneity**

**Table 2 Real parts of parameter $F_k^{(3)}$ in the case of "b-deffect" in one cell**

| Cell N | | $b_{25} = b_{25}^{(0)} - 0.002cm$ | $b_{22} = b_{22}^{(0)} - 0.002cm$ | $b_{28} = b_{28}^{(0)} - 0.002cm$ | $b_{23} = b_{23}^{(0)} - 0.002cm$ |
|---|---|---|---|---|---|
| 21 | -2.00E-15 | -3.00E-09 | 6.69E-05 | -3.00E-09 | 8.88E-09 |
| 22 | 1.33E-15 | 1.11E-05 | **-9.68E-02** | 1.11E-05 | -7.62E-05 |
| **23** | **6.66E-16** | **-3.76E-06** | **8.05E-05** | **-3.75E-06** | **1.03E-01** |
| **24** | **-9.99E-16** | **7.84E-05** | **-8.00E-09** | **7.12E-06** | **-6.97E-05** |
| 25 | -2.22E-16 | **-9.68E-02** | 6.42E-10 | -1.11E-05 | 7.19E-09 |
| 26 | 8.88E-16 | 7.56E-05 | -5.02E-13 | -5.83E-09 | -6.31E-10 |
| 27 | -2.22E-16 | -7.07E-09 | 2.22E-16 | 6.69E-05 | 4.97E-13 |
| 28 | 2.22E-16 | 6.60E-10 | 4.44E-16 | **-9.68E-02** | 4.44E-16 |
| 29 | 2.22E-16 | -5.72E-13 | -2.22E-16 | 7.56E-05 | 0.00E+00 |

**Table 3 Real parts of parameter $F_k^{(3,00)}$ in the case of "b-deffect" in one cell**

| Cell N | | $b_{25}=b_{25}^{(0)}-0.002cm$ | $b_{22}=b_{22}^{(0)}-0.002cm$ | $b_{28}=b_{28}^{(0)}-0.002cm$ | $b_{23}=b_{23}^{(0)}-0.002cm$ |
|---|---|---|---|---|---|
| 21 | -1.39E-04 | -1.39E-04 | -7.16E-05 | -1.39E-04 | -1.39E-04 |
| 22 | -3.99E-05 | -1.77E-05 | **-9.68E-02** | -1.77E-05 | -1.11E-04 |
| **23** | **-4.81E-02** | **-4.90E-02** | **-4.80E-02** | **-4.90E-02** | **4.87E-02** |
| **24** | **-4.81E-02** | **-4.63E-02** | **-4.81E-02** | **-4.64E-02** | **-4.82E-02** |
| 25 | -3.98E-05 | **-9.69E-02** | -3.98E-05 | -6.20E-05 | -3.98E-05 |
| 26 | -1.39E-04 | -6.30E-05 | -1.39E-04 | -1.39E-04 | -1.39E-04 |
| 27 | -1.39E-04 | -1.39E-04 | -1.39E-04 | -7.16E-05 | -1.39E-04 |
| 28 | -1.39E-04 | -1.39E-04 | -1.39E-04 | **-9.69E-02** | -1.39E-04 |
| 29 | -1.39E-04 | -1.39E-04 | -1.39E-04 | -6.30E-05 | -1.39E-04 |

**Table 4 Real parts of parameter $F_k^{(3,0)}$ in the case of "b-deffect" in one cell**

| Cell N | | $b_{25}=b_{25}^{(0)}-0.002cm$ | $b_{22}=b_{22}^{(0)}-0.002cm$ | $b_{28}=b_{28}^{(0)}-0.002cm$ | $b_{23}=b_{23}^{(0)}-0.002cm$ |
|---|---|---|---|---|---|
| 21 | -1.38E-04 | -1.38E-04 | **2.22E-03** | -1.38E-04 | -1.39E-04 |
| 22 | **2.64E-03** | **3.03E-03** | **-9.88E-02** | **3.03E-03** | -1.14E-04 |
| **23** | **-5.13E-02** | **-5.24E-02** | **-4.87E-02** | **-5.24E-02** | **5.05E-02** |
| **24** | **-4.97E-02** | **-4.53E-02** | **-4.97E-02** | **-4.77E-02** | **-5.21E-02** |
| 25 | **2.07E-03** | **-9.97E-02** | **2.07E-03** | **1.80E-03** | 2.07E-03 |
| 26 | -1.38E-04 | **2.52E-03** | -1.38E-04 | -1.37E-04 | -1.38E-04 |
| 27 | -1.39E-04 | -1.38E-04 | -1.39E-04 | **2.21E-03** | -1.39E-04 |
| 28 | -1.39E-04 | -1.39E-04 | -1.39E-04 | **-0.10159** | -1.39E-04 |
| 29 | -1.39E-04 | -1.39E-04 | -1.39E-04 | **2.52E-03** | -1.39E-04 |

Possibility of using more simple coefficients[9] $F_k^{(3,00)}, F_k^{(3,0)}$ (especially $F_k^{(3,0)}$) as criterion of cell tuning are illustrated by Table 3 and Table 4. From these tables it follows that the influence of the defects in neighboring cells on the values of coefficients $F_k^{(3,00)}, F_k^{(3,0)}$ is measured in a few thousandths. For evaluation of such influence we made calculation for the case when the inhomogeneity under consideration is slightly detuned (see Table 5). Introduced defects resulted in $F_k^{(3,0)}$ value change about 4E-3 (the reflection coefficient equals 2E-3, for matched conditions it equals 1.5E-4). Distributions of amplitudes and phases for this case are presented in Fig. 22 and Fig. 23. We can see that the amplitude changes are negligible and the phase changes are about a few tenths of degree.

**Table 5 Real parts of parameter $F_k^{(3,0)}$ in the case of "b-defect" in the cells N=23,24**

| Cell N | | $b_{24}=b_{23}=b_{23}^{(0)}+0.000075\,cm$ |
|---|---|---|
| 21 | -1.38E-04 | -1.38E-04 |
| 22 | 2.64E-03 | 2.73E-03 |
| 23 | **-5.13E-02** | **-5.51E-02** |
| 24 | **-4.97E-02** | **-5.36E-02** |
| 25 | 2.07E-03 | 2.18E-03 |
| 26 | -1.38E-04 | -1.38E-04 |

[9] Nonzero values is not the obstacle for using these coefficients. We can calculate them during the simulation process.

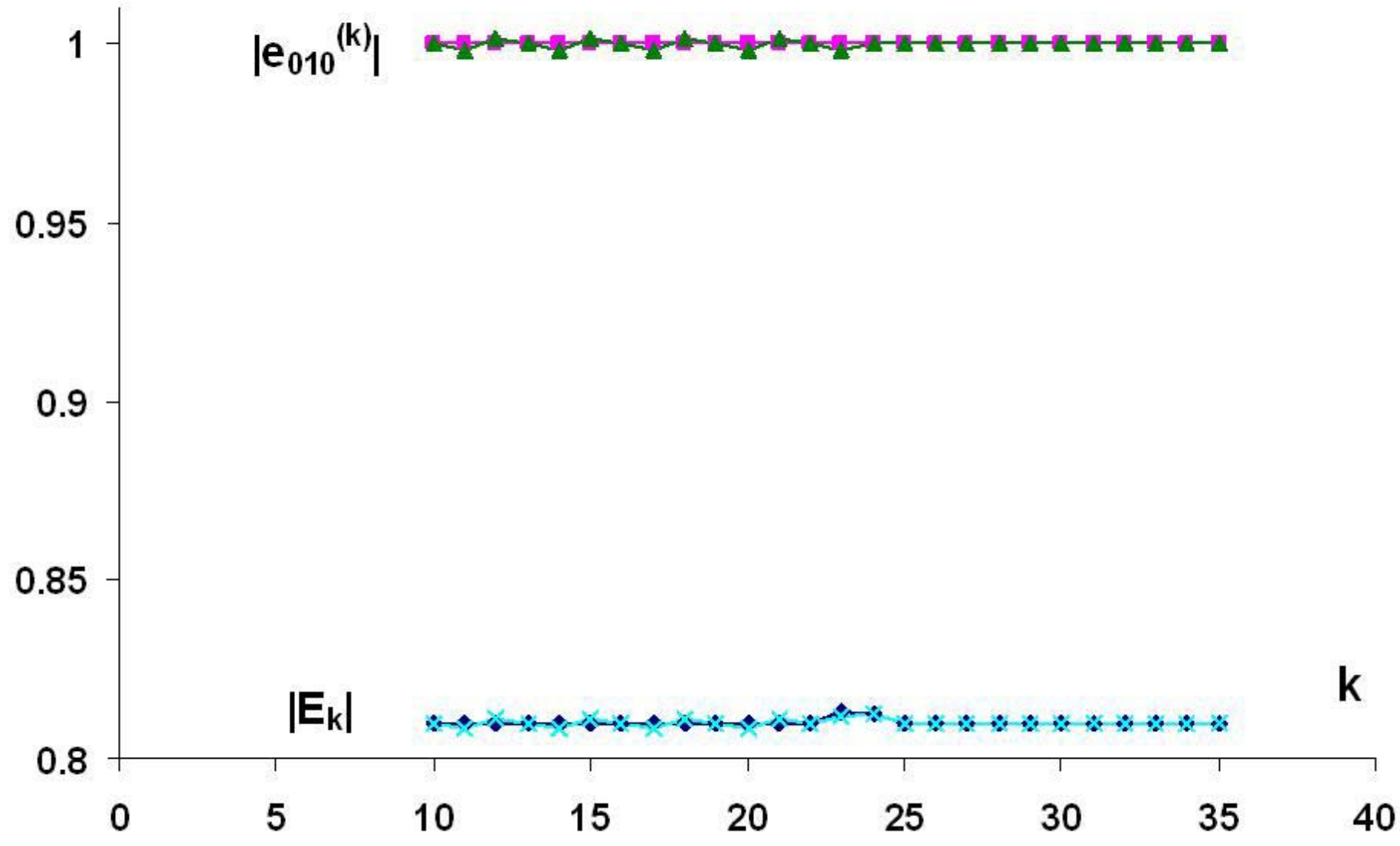

**Fig. 22 Modules amplitudes of $E_{010}$ mode and fields in the centers of cells with and without defects**

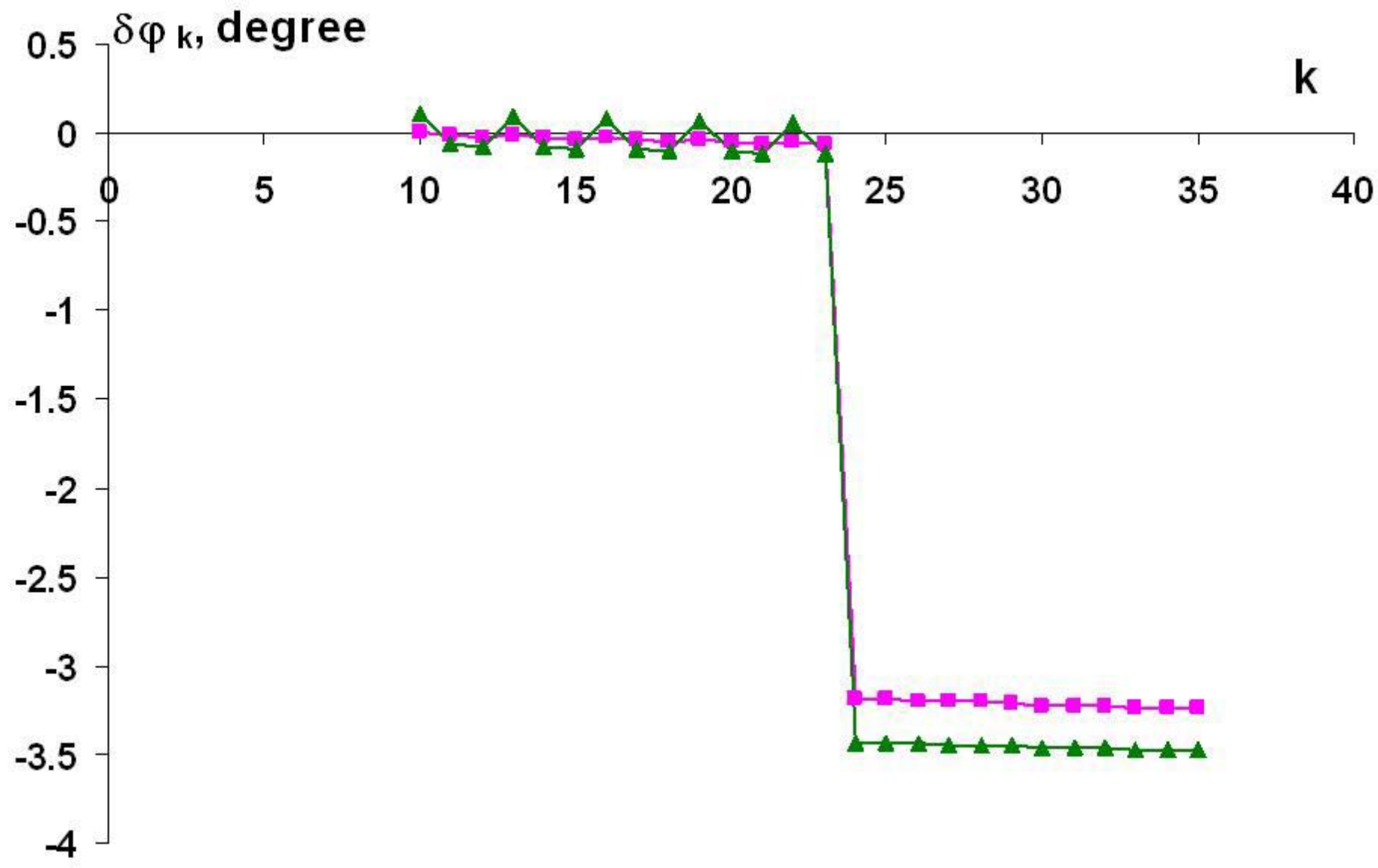

**Fig. 23 Phase shifts ($\varphi_k - 120^\circ$) in the centers of cells with and without defects**

These results show that one can use coefficients $F_k^{(3,0)}$ under tuning DLWs with small inhomogeneity and coupling (the phase velocity is close to the velocity of light and radii of holes are not large). But before the starting of tuning process it is needed to calculate the necessary values of these coefficients[10] and use them under tuning. If diaphragms are produced with good quality, the accuracy of such tuning can be a few tenths of degree.

If one will use the zero value for coefficients $F_k^{(3,0)}$ of the tuned DLW [11-19], the errors of tuning becomes greater. Fig. 24, Fig. 25 and Fig. 26 illustrate this case. We can see that accuracy of such tuning can be a few degrees in phase shifts. There is also amplitude oscillation along structure.

[10] For simplest case of the uniform DLW these coefficients are constant and have small values (see above).

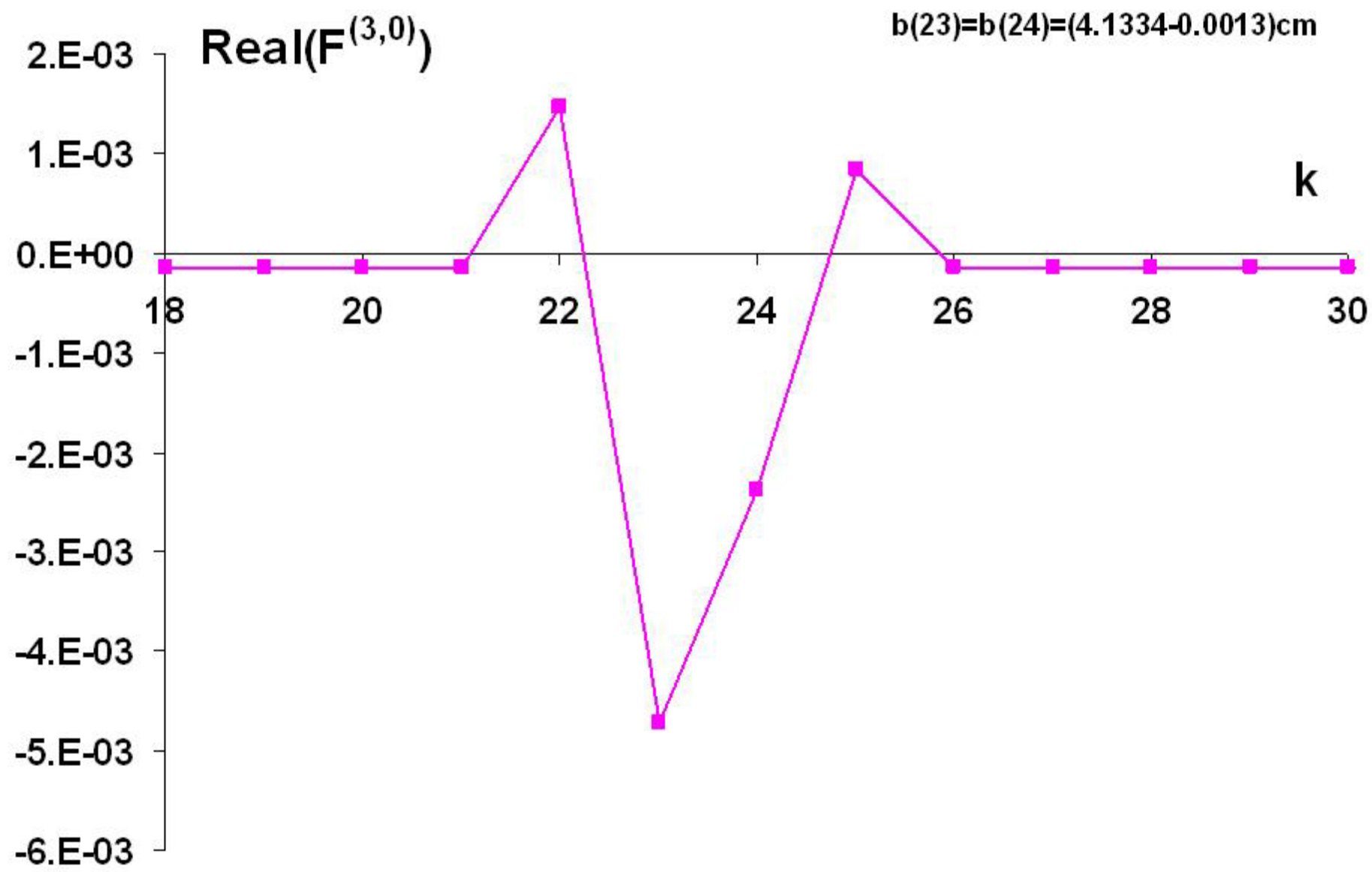

**Fig. 24 Real parts of parameter** $F_k^{(3,0)}$ **in the centers of cells for the case when** $F_k^{(3,0)} \to 0$

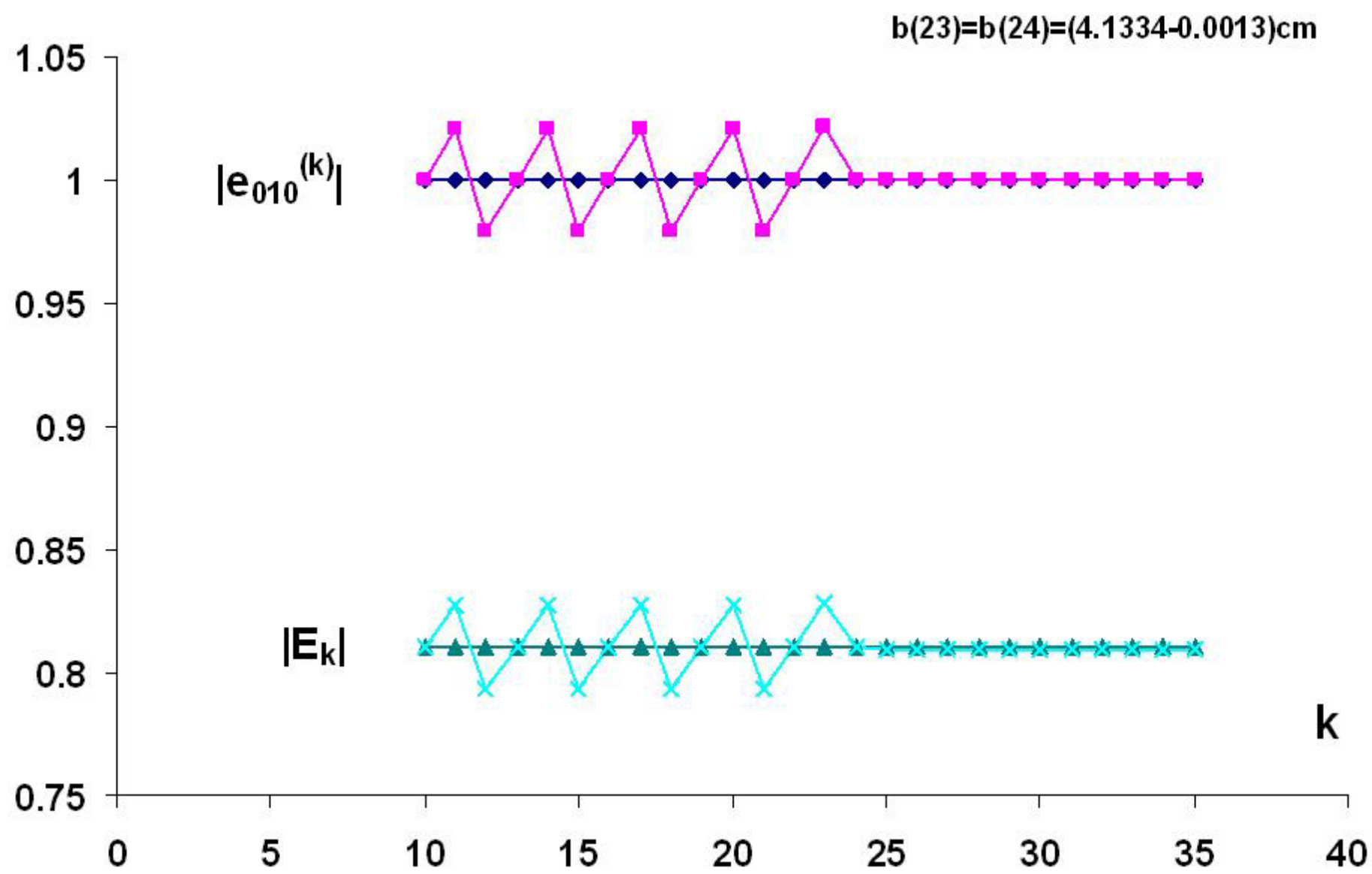

**Fig. 25 in the centers of cells for the case when** $F_k^{(3,0)} \to 0$

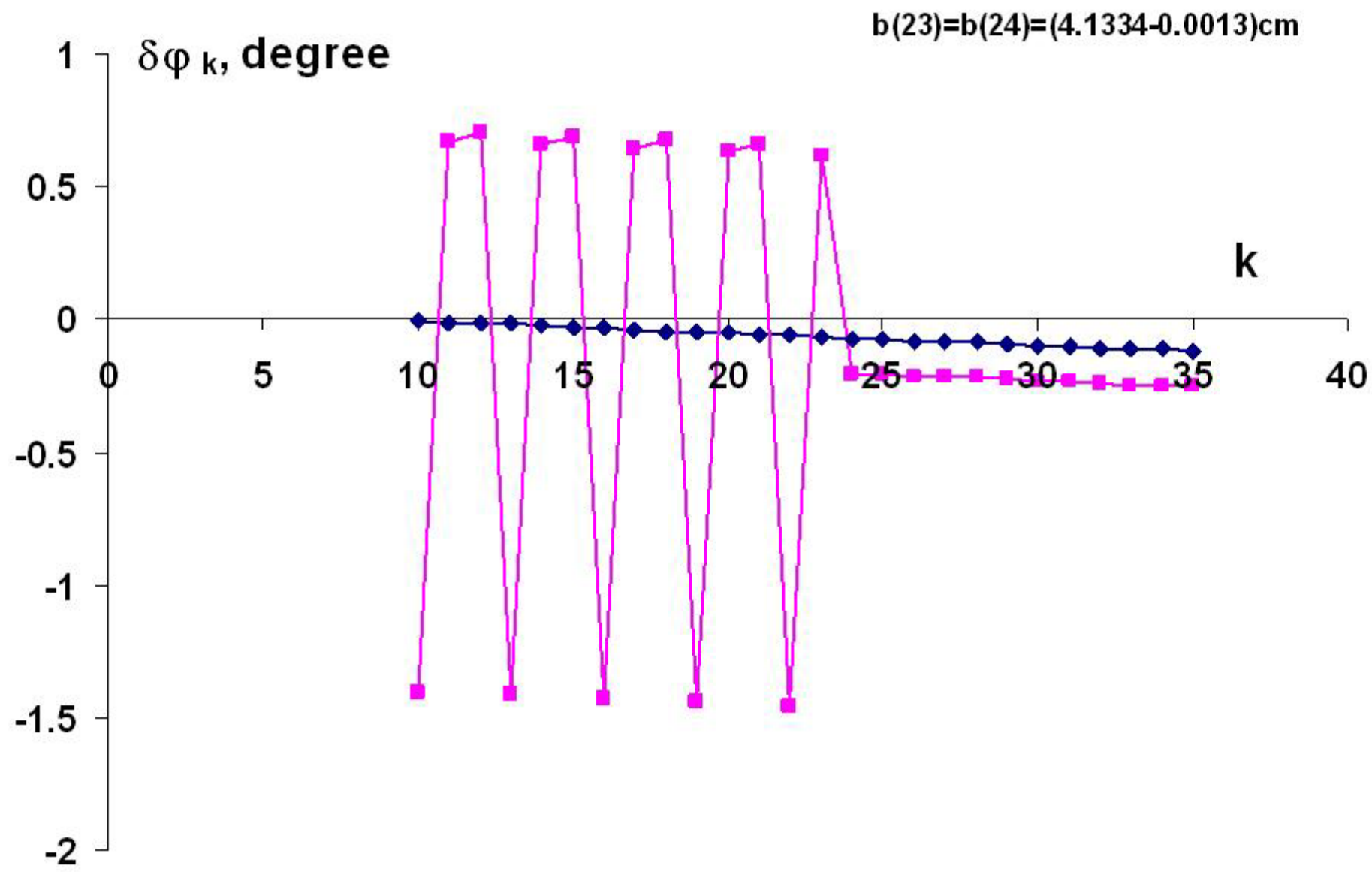

**Fig. 26 Phase shifts ($\varphi_k - 120^\circ$) in the centers of cells for the case when $F_k^{(3,0)} \to 0$**

## Conclusions

Results of conducted investigations show that proposed mode matching technique can be effectively used for calculation of parameters of nonuniform DLWs. Some of these parameters can be used for tuning nonuniform DLWs with the field measurement method. Accuracy of this method under tuning of DLWs with $\beta_{ph} \sim 1$ is evaluated. As it follows from our investigation for DLWs with $\beta_{ph} \sim 1$ one can use simple coefficients in tuning process, but the needed values of these coefficients must be obtained by calculation of DLW parameters.

## References

1 S.K.Katenev Eigenwave characteristics of a periodic iris-loaded circular waveguide The concepts. PIER, 2007, V.69, pp.177-200

2 L.F. Wang, Y. Z. Lin, T. Higo et al. High accuracy analysis of arbitrary modes in tapered disk-loaded structures. Proceedings of PAC2001, pp.3045-3047

3 P.M. Lapostolle, A.L. Septier, Linear accelerators, Amsterdam: North-Holland Publ. Co., 1970

4 T.P. Wangler, RF linear accelerators, Wiley-VCH, 2008

5 S.E.Tsimring Electron Beams and Microwave Vacuum Electronics John Wiley & Sons, 2007

6 M.I.Ayzatsky. New Mathematical Model of an Infinite Cavity Chain. Proceedings of the EPAC96, 1996,v.3, p.2026-2028; On the Problem of the Coupled Cavity Chain Characteristic Calculations. http://xxx.lanl.gov/pdf/acc-phys/9603001.pdf, LANL.arXiv.org e-print archives, 1996.

7 M.Ayzatskiy, K.Kramarenko. Coupling coefficients in inhomogeneous cavity chain Proceedings of the EPAC2004, 2004, pp.2759-2761;

8 M.I. Ayzatskiy, V.V. Mytrochenko. Coupled cavity model based on the mode matching technique. http://lanl.arxiv.org/ftp/arxiv/papers/1505/1505.03223.pdf, LANL.arXiv.org e-print archives, 2015.

9 M.I. Ayzatskiy, V.V. Mytrochenko. Electromagnetic fields in nonuniform disk-loaded waveguides. http://lanl.arxiv.org/ftp/arxiv/papers/1505/1505.03223.pdf, , LANL.arXiv.org e-print archives, 2015 10 N.I.Aizatsky. On the Theory of Two Coupled Cavities. Proceedings of the PAC95, 1995, v.3, p.1773-1775; On two cavity coupling, arXiv:acc-phys/9603002, LANL.arXiv.org e-print archives 1996

11 T.Khabiboulline, V.Puntus, M.Dohlus et al. A new tuning method for traveling wave structures. Proceedings of PAC95, pp.1666-1668

12 Jiaru Shi, Alexej Grudiev, Andrey Olyunin, Walter Wuensch. Tuning of CLIC accelerating structure prototypes at CERN. Proceedings of LINAC2010, pp.97-99

13 Fang WenCheng, Tong DeChun, Gu Qiang, Zhao ZhenTang. Design and experimental study of a C-band traveling-wave accelerating structure. Chinese Science Bulletin, 2011, v.56, N.1, pp.18-23

14 W. C. Fang, Q. Gu, Z. T. Zhao, D. C. Tong. The nonresonant perturbation theory based field measurement and tuning of a linac accelerating structure. Proceedings of LINAC2012, pp.375-377

15 Ang WenCheng, Tong DeChun, Gu Qiang, Zhao ZhenTang, Sheng Xing, Chen LiFang, Wang Lin.The nonresonant perturbation theory based field measurement and tuning of a linac accelerating structure. Science China, 2013, Vol.56, N.11, pp.2104–2109

16 Gong Cun-Kui. Zheng Shu-Xin, Shao Jia-Hang, JIA Xiao-Yu, Chen Huai-Bi. A tuning method for nonuniform traveling-wave accelerating structures. Chinese Physics C, 2013, V.37, N.1, 017003

17 J.Shi, A.Grudiev, W.Wuensch. Tuning of X-band traveling-wave accelerating structures, Nuclear Instruments and Methods in Physics Research, 2013, A704,pp.14–18

18 Y.C.Nien, C.Liebig,M.Hüning,M.Schmitz. Tuning of 2.998 GHz S-band hybrid buncher for injector upgrade of LINAC II at DESY. Nuclear Instruments and Methods in Physics Research 2014, A761, pp.69–78

19 R. Wegner, W. Wuensch, G. Burt, B. Woolley. Bead-pull measurement method and tuning of a prototype CLIC CRAB cavity, Proceedings of LINAC2014, pp.134-136

20 R.B Neal, General Editor, The Stanford Two-Mile Accelerator, New York, W.A. Benjamin, 1968